\documentclass{aa}  

\usepackage{graphicx}
\usepackage{txfonts}
\usepackage{natbib}
\usepackage{hyperref}
\usepackage{subcaption}
\begin{document}

   \title{Features and prospects for kilonova remnant detection with current and future surveys}


   \author{Sandeep Kumar Acharya\thanks{Corresponding author}
          \inst{1},
          Paz Beniamini
          \inst{1,2,3}
          and 
          Kenta Hotokezaka
          \inst{4}
          }

   \institute{Astrophysics Research Center of the Open University, The Open University of Israel, Ra'anana, Israel\\
              \email{sandeepkumaracharya92@gmail.com}
         \and
             Department of Natural Sciences, The Open University of Israel, P.O Box 808, Ra’anana 4353701, Israel
          \and
             Department of Physics, The George Washington University, 725 21st Street NW, Washington, DC 20052, USA
           \and  Research Center for the Early Universe, Graduate School of Science, The University of Tokyo, Tokyo 113-0033, Japan
             }

   \date{}

 
 \abstract{We study the observable spectral and temporal properties of kilonova remnants (KNRs) analytically, and point out quantitative differences with respect to supernova remnants. We provide detection prospects of KNRs in the context of ongoing radio surveys. We find that there is a good chance to expect tens of these objects in future surveys with a flux threshold of $\sim 0.1$ mJy. Kilonova remnants from a postulated population of long-lived supermassive neutron star remnants of neutron star mergers are even more likely to be detected, as they are extremely bright and peak earlier. For an ongoing survey with a threshold of $\sim$ 1 mJy, we expect to find tens to hundreds of such objects if they are a significant fraction of the total kilonova (KN) population. Considering that there are no such promising KN candidates in presently ongoing surveys, we constrain the fraction of these extreme KN to be no more than 30 percent of the overall KN population. This constraint depends sensitively on the details of ejecta mass and external density distribution.  }


\titlerunning{detection prospects of kilonovae remnants}
\authorrunning{Acharya et. al.}

   \maketitle
   
%

\section{Introduction}
Kilonovae (KNe) are transient events formed after the merger of binary neutron stars \citep{1998ApJ...507L..59L,2010MNRAS.406.2650M,2015IJMPD..2430012R,2016AdAst2016E...8T,2016ARNPS..66...23F,HB2018,2019LRR....23....1M}. Their emission initially peaks in the UV-optical-IR range and arises from heated ejecta material. The heat is generated from radioactive decay of heavy elements produced through r-process nucleosynthesis. The ejecta mass is expected to be in the range of $0.01-0.1M_{\odot}$, with a total kinetic energy of $\sim 10^{51}$ erg. If the merger results in a long-lived rapidly spinning and highly magnetized neutron star (millisecond magnetar), then a large amount of energy (typically $\sim 5\times 10^{52}$\,erg up to $\sim 2\times 10^{53}$\,erg --- limited primarily by the maximum possible rotation that the neutron star can attain before being torn apart, which is known as the mass-shedding limit; see, e.g. \citealt{2013ApJ...771L..26G,2021ApJ...920..109B}) can be injected into the outflow. This can significantly enhance the brightness of the resulting KN \citep{2013ApJ...776L..40Y,2014MNRAS.439.3916M,2016ApJ...818..104K,2022MNRAS.516.2614A,2024MNRAS.527.5166W}. 

In association with the gravitational-wave-detected neutron star merger,  GW170817 \citep{Abott2017}, a KN with ejecta  of much lower kinetic energy, of the order of $\sim 10^{51}$ erg, has been observed with optical follow-up studies \cite{Coulter2017,Soares2017,Arcavi2017,Lipunov2017,Valenti2017,Troja2017,Kilpatrick2017,Smartt2017,Drout2017,Evans2017}, suggesting a long-lived magnetar was not formed in this event. 
While supermassive neutron star (NS) formation might have been thought to be common based on considerations of total mass alone (e.g. \citealt{2019ApJ...880L..15M}), the occurrence rate of magnetar-boosted KNe (and the underlying long-lived central engines) is in practice strongly constrained by inferences based on observations. Relevant examples are (i) the analysis in \cite{2021ApJ...920..109B}, which is based on the typical durations and ejected energies of KN central engines as compared with the plateau-like features in short gamma-ray burst (sGRB) gamma-ray and X-ray observations; (ii) the long-term radio follow-up of low-redshift sGRBs (e.g. \citealt{2016ApJ...819L..22H,Schroeder2020,2021MNRAS.500.1708R}), which puts limits on the kinetic energy of the outflow that are well below values expected for magnetar-boosted KNe (excluding the maximum ejecta energy of $\gtrsim 5\times 10^{52}$ erg for $M_{\rm ej}\lesssim 0.05M_{\odot}$, and constraining the fraction of short GRBs that produce stable remnants to $\lesssim 0.5$);  and (iii) the lack of bright transients in optical all-sky surveys such as \textit{Zwicky} Transient Facility \citep[ZTF;][]{ZTF2019}, which would naturally arise from magnetar-boosted KNe \citep{2024MNRAS.527.5166W}. The fraction of magnetar-boosted KNe is important for constraining the NS equation of state, because the latter is a major factor in determining the outcome of a NS merger \citep{Shibata2019}. 

While the optical transient from a KN substantially fades after a week or two, the ejecta continues to expand into the surrounding medium. As a result, a forward shock is driven into the surrounding material. This shock enhances the magnetic fields and accelerates electrons in the shocked medium, which then radiate their energy via synchrotron and synchrotron self-Compton. This can give rise to emission that is sometimes known as the `kilonova afterglow' and can observationally manifest as approximately decades-long radio flares \citep{2011Natur.478...82N,2014PhRvD..89f3006T,
2015MNRAS.452.3419M,2015MNRAS.450.1430H,2019MNRAS.487.3914K}. In the case of GW170817, the KN afterglow has not yet been observed even 7 years after the merger (e.g. \citealt{2019ApJ...886L..17H,2022ApJ...938...12B}). 

Since the KN ejecta blast waves are expected to be mildly relativistic at most, the physics is essentially the same as in supernova remnants (SNRs). In analogy with the latter, this phase might also be dubbed the kilonova remnant \citep[KNR;][]{2021ApJ...920..109B}. Indeed, at times later than the `deceleration peak' (also known as the Sedov time), the evolution depends only on the energy of the blastwave, and as such ---considering the similarity between the energies of supernovae (SNe) and KNe---,  it  becomes challenging to distinguish between the two remnant types. The problem is compounded by the fact that the occurrence rate of SNe is about $ 300$ times greater than that of KNe, and so SNRs are the major contaminant preventing potential identification of KNRs. Nevertheless, we show in the present work that there are important spectral, temporal, and environmental properties that can be used to distinguish KNRs from SNRs in observations.  

In this paper, we perform an analytic study of the expected observable properties of KNRs and highlight their main differences from SNRs. In addition, we also focus on the class of highly energetic magnetar-boosted KN remnants (EKNRs), which, though expected to be rare, can be seen to much greater distances; therefore even a null detection of KNRs could strongly constrain their intrinsic occurrence rate.
This paper is organised as follows. We start by presenting the dynamics of the ejecta of KNe in Sect. \ref{sec:dynamics_newtonian}. We use Newtonian dynamics, which is relevant for our choice of fiducial parameters. However, we add a brief section (Appendix \ref{subsec:dynamics_relativistic}) with fully relativistic expressions. In Sect. \ref{sec:electron_properties}, we obtain expressions for the flux from emission originating from non-thermal electrons. We describe the combined spectrum due to synchrotron, bremsstrahlung, and synchrotron self-Compton emission and emphasise the differences between SNRs and KNRs. We predict the number of expected KNRs in radio surveys in Sect. \ref{sec:N_remnants} and show that we are likely to detect a candidate object with current (or future) facilities such as VLASS (Very Large Area Sky Survey)  and VAST (Variables and Slow Transients). We are even more likely to detect an extreme KN, if they exist, due to the strong dependency of the observed flux on ejecta energy. As of yet, no such candidate object has been reported in the literature, which allows us to put moderate constraints on their abundance. In Sect. \ref{sec:detection_prospects}, we discuss how to extract the Sedov time from its observed light curve. For EKNRs, the Sedov time is of the order of 10-20 years, and therefore there is a good chance to obtain the Sedov time of extreme KNs with dedicated observations over a period of a few years. We also discuss how information about the host galaxies can be used to break degeneracies between KNRs and SNRs.
We conclude in Sect. \ref{sec:conclusions}.
\section{Ejecta dynamics}
\label{sec:dynamics_newtonian}
We compute the evolution of an ejecta (associated with either a KN or a SN) assuming it expands into a uniform surrounding medium in a spherically symmetric fashion. We assume Newtonian physics for the calculations shown here (see Sect. \ref{subsec:dynamics_relativistic} for a fully relativistic treatment). The ejecta's kinetic energy and velocity are connected via
\begin{equation}
    E=\frac{1}{2}(M_{\rm ej}+M_{\rm ISM})v_{\rm ej}^2,
    \label{eq:KN_velocity}
\end{equation}
where $M_{\rm ej}$ is the ejecta mass and $M_{\rm ISM}$ is the mass of displaced surrounding gas with $M_{\rm ISM}=\frac{4\pi}{3}m_{\rm p}n_{\rm ISM}R_{\rm ej}^3$, $m_{\rm p}$ is the the mass of proton, $n_{\rm ISM}$ is the density of the surrounding gas and $R_{\rm ej}$ is the size of expanding ejecta. The kinetic and the internal energy of the ejecta can be comparable. Therefore, if we were to use the total energy of ejecta in Eq. \ref{eq:KN_velocity}, it would roughly cancel the factor 1/2 on the right-hand side. 
We assume a uniform density profile without any spatial gradient. We normalise the physical parameters by defining, $E_{51}=\frac{E\hspace{0.1cm} }{10^{51} {\rm erg}}, M_{\rm ej,\odot}=\frac{M_{\rm ej}}{M_{\odot}}$ , and $n_0=\frac{n_{\rm ISM}}{\rm 1 cm^{-3}}$. During the initial phase, the ejecta expands freely into the medium (we call this the `coasting phase' regime). This lasts until $M_{\rm ej}\sim M_{\rm ISM}$, when the ejecta starts decelerating and the evolution enters the Sedov-Taylor (ST) phase. The deceleration, or ST radius, $R_{\rm ST}$ can be obtained by solving
\begin{equation}
    \frac{E}{2}=\frac{4\pi}{3}n_{\rm ISM}R_{\rm ST}^3m_{\rm p}{\rm c^2}\frac{\beta_{\rm ej}^2}{2},
\end{equation}
where $v_{\rm ej}=\beta_{\rm ej}c$. This simplifies to $R_{\rm ST}=0.52n_0^{-1/3} M_{\rm ej,\odot}^{1/3}\times 10^{19} {\rm cm}$. 
The corresponding timescale is given by
\begin{equation}
    t_{\rm ST}=165n_0^{-1/3}E_{51}^{-1/2}M_{\rm ej,\odot}^{5/6} {\rm  yr.}
    \label{eq:t_ST}
\end{equation}
The velocity in each regime is
\begin{equation}
  \begin{aligned}
    \beta_{\rm ej}=\frac{v_{\rm ej}}{\rm c}=0.03\left(\frac{E_{51}}{M_{\rm ej,\odot}}\right)^{1/2}\hspace{0.1cm} ({\rm coasting\hspace{0.1cm} phase}) \\
    \beta_{\rm ej}=0.52E_{51}^{1/5}n_0^{-1/5}t_{\rm yr}^{-3/5}\hspace{1cm} ({\rm ST\hspace{0.1cm} phase}).
    \end{aligned}
    \label{eq:beta_ej}
\end{equation}
The radius of the ejecta is then $R_{\rm ej}\approx \beta_{\rm ej} c t$ in the coasting phase and $R_{\rm ej}\approx \frac{5}{2}\beta_{\rm ej} c t$ in the ST phase.
We consider three classes of objects with fiducial physical parameters as stated in parentheses: a SNR ($E_{51}=$ 1, $M_{\rm ej,\odot}$= 1, $n_0$= 1), a KNR ($E_{51}=1$, $M_{\rm ej,\odot}$= 0.1, $n_0$= 0.1), and an EKNR ($E_{51}=10$, $M_{\rm ej,\odot}$= 0.1, $n_0$= 0.1). We note that this implicitly assumes that, compared to SNRs, both KNRs and EKNRs occur in environments  that exhibit less star formation and are therefore less dense. While this is true on average, it means that KNRs and EKNRs appear less bright than they would in a similar environment to a SNR, which is a topic we return to in Sect. {\ref{subsec:KNR_vs_SNR}}.

In these calculations, we have assumed that the external medium is undisturbed until the ejecta collides with it. However, relativistic jets from merging binaries can disturb the external medium before the ejecta. In that case, the early phase of the KNR emission could be suppressed due to the blast-wave passing through a region that was evacuated by the passage of the jet \citep{2020MNRAS.495.4981M}. The timescale of this suppression is related to $t_{\rm ST}$ by $t_{\rm col}=2.04\left(\frac{E_{\rm j}}{E}\right)^{1/3}t_{\rm ST}$.
For a jet with an isotropic equivalent energy of $10^{51}$ erg \citep{DG2015} and an opening angle of $\theta_j\sim 0.2$ radians, the energy of the jet $E_{\rm j}=\frac{\theta_j^2}{2}E_{\rm iso}\sim 10^{49}$ erg. For this case, we find that the suppression timescale is significantly shorter that the Sedov timescale. This means that propagation through the evacuated region typically only changes the KNR flux at early times, when the KNR is generally very dim. This conclusion is further enhanced in the case of EKNRs.
\section{Synchrotron, synchrotron self-Compton, and Bremsstrahlung emission}
\label{sec:electron_properties}
The expanding ejecta shocks the surrounding medium, compressing it and accelerating the electrons. We assume that a fraction $\epsilon_e$ of the shock energy goes into accelerating electrons and a fraction $\xi_e$ of electrons are accelerated to relativistic energies.

The relativistic electron Lorentz factor distribution is taken to be of the form
\begin{equation}
    \frac{{\rm d}N_e}{{\rm d}\gamma_e}(\gamma_e)=\frac{N_0}{\gamma_m} \left(\frac{\gamma_e}{\gamma_m}\right)^{-p},
\end{equation}
where $\gamma_m$ is the minimal Lorentz factor. Integrating over $\gamma_e$ and equating the electron number density to $n_{\rm ISM}$, $N_0=(p-1)n_0$.
The expression for $\gamma_m$ in the bulk comoving frame is given by
\begin{equation}
\label{eq:gm1}
    \gamma_m-1=\frac{p-2}{p-1}\frac{\epsilon_e}{\xi_e} \frac{m_{\rm p}}{m_{\rm e}}(\Gamma-1),
\end{equation}
where it is assumed that $p>2$ with a fiducial value of $p=2.5$ and $\Gamma$ is the Lorentz factor of the shocked fluid. In the non-relativistic regime of expanding ejecta and defining $\bar{\epsilon_e}=\epsilon_e\frac{p-2}{p-1}$, $\bar{\epsilon_e}=0.1\bar{\epsilon}_{e,-1}$, the expression simplifies to, $\gamma_m-1=91.85\frac{\bar{\epsilon}_{e,-1}}{\xi_e}\beta_{\rm ej}^2$, or
\begin{equation}
    \begin{aligned}
    \label{eq:gmtwocases}
       \gamma_m-1 &= 8.2\times 10^{-2} \frac{\bar{\epsilon}_{e,-1}}{\xi_e}\left(\frac{E_{\rm 51}}{M_{\rm ej,\odot}}\right) \hspace{0.5cm} {(\rm coasting\hspace{0.1cm}phase)} \\
     \gamma_m-1 &= 45.9\frac{\bar{\epsilon}_{e,-1}}{\xi_e}E_{51}^{2/5}n_0^{-2/5}t_{\rm yr}^{-6/5}\hspace{0.8cm} {(\rm ST\hspace{0.1cm}phase)}.
    \end{aligned}
\end{equation}
For a fixed $\xi_e$ and at sufficiently late times, Eq. \ref{eq:gmtwocases} would lead to $\gamma_m-1\ll 1$, which will result in electrons being non-relativistic, at which point they cannot emit synchrotron photons at $\nu\gtrsim$ GHz. The reason for this is straightforward: at such times, there is simply not enough energy to share between those electrons, while still allowing all of them to maintain relativistic velocities. This phase is known as the deep Newtonian phase. A natural resolution, which is consistent with giant flare outflows and GRB afterglows \citep{Granot2006,SG2013}, is to take $\gamma_m$ to be the maximum between the value given in Eq. \ref{eq:gmtwocases} and $\sqrt{2}$. The fraction of relativistic electrons then needs to be modified according to $\xi_e=\xi_{e,0}{\rm min}[1,(\beta_{\rm ej}/\beta_{\rm dn})^2]$ \citep{BGG2022}.
In the analysis below, we assume $\xi_{e,0}=1$.
The expression for $\beta_{\rm dn}$ is then $\beta_{\rm dn}=\left(2^{3/2}\frac{p-1}{p-2}\frac{10}{\bar{\epsilon}_{e,-1}}\frac{m_{\rm e}}{m_{\rm p}}\right)^{1/2}$.
Using $p=2.5$ and $\bar{\epsilon}_{e,-1}=1$, $\beta_{\rm dn}=0.21$. For our fiducial class of objects, we find that only the EKNRs transition to the deep Newtonian regime, while SNRs and KNRs are already in this regime in the coasting phase. As the ejecta transitions to the deep Newtonian regime in the ST era, requiring $\beta_{\rm ej}=\beta_{\rm dn}$ in Eq. \ref{eq:beta_ej}, we obtain the transition timescale as,
\begin{equation}
    t_{\rm dn}=7.43\left(\bar{\epsilon}_{e,-1}^{1/2}E_{51}^{1/5}n_0^{-1/5}\right)^{5/3}\hspace{0.2cm} {\rm yr}.
\end{equation}
For $E_{51}=10, n_0=0.1,$ and $\bar{\epsilon}_{e,-1}=1$, $t_{\rm dn}\approx 30$ yr. The expression for
$\xi_e$ is given by
\begin{equation}
\begin{aligned}
    \xi_e\sim 2.25\times 10^{-2}\frac{E_{\rm 51}}{M_{\rm ej,\odot}} {\bar{\epsilon}_{e,-1}}\hspace{0.2cm} {(\rm coasting\hspace{0.1cm}phase)} \\
    \xi_e\sim 6.3E_{\rm 51}^{2/5}{\bar{\epsilon}_{e,-1}}n_0^{-2/5}t_{\rm yr}^{-6/5}\hspace{0.8cm} {(\rm ST\hspace{0.1cm}phase)}.
    \end{aligned} 
    \label{eq:xie}
\end{equation}
\subsection{Synchrotron emission}
\label{sec:synchrotron}
\begin{figure*}
\centering 
\includegraphics[width=\columnwidth]{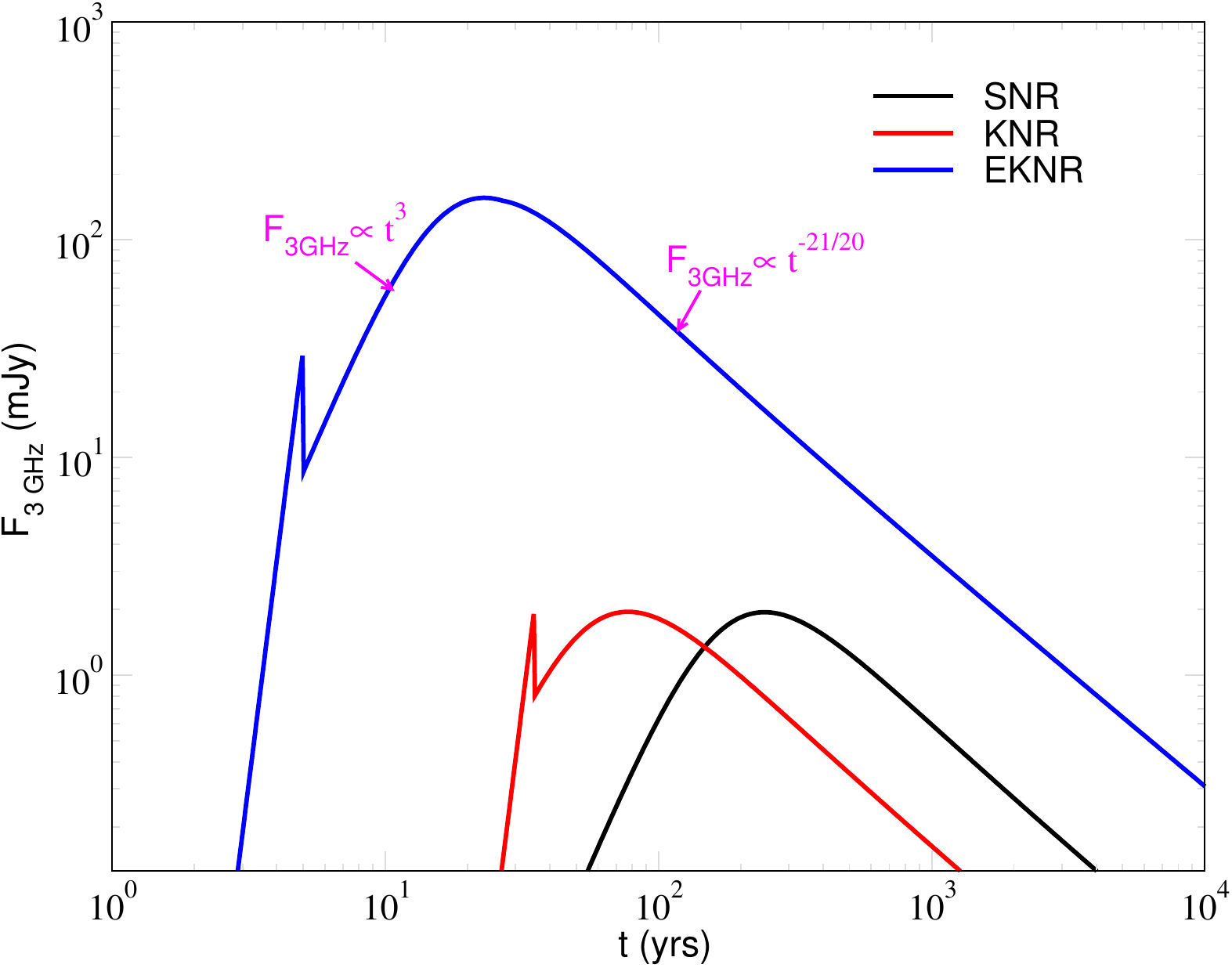}
\includegraphics[width=\columnwidth]{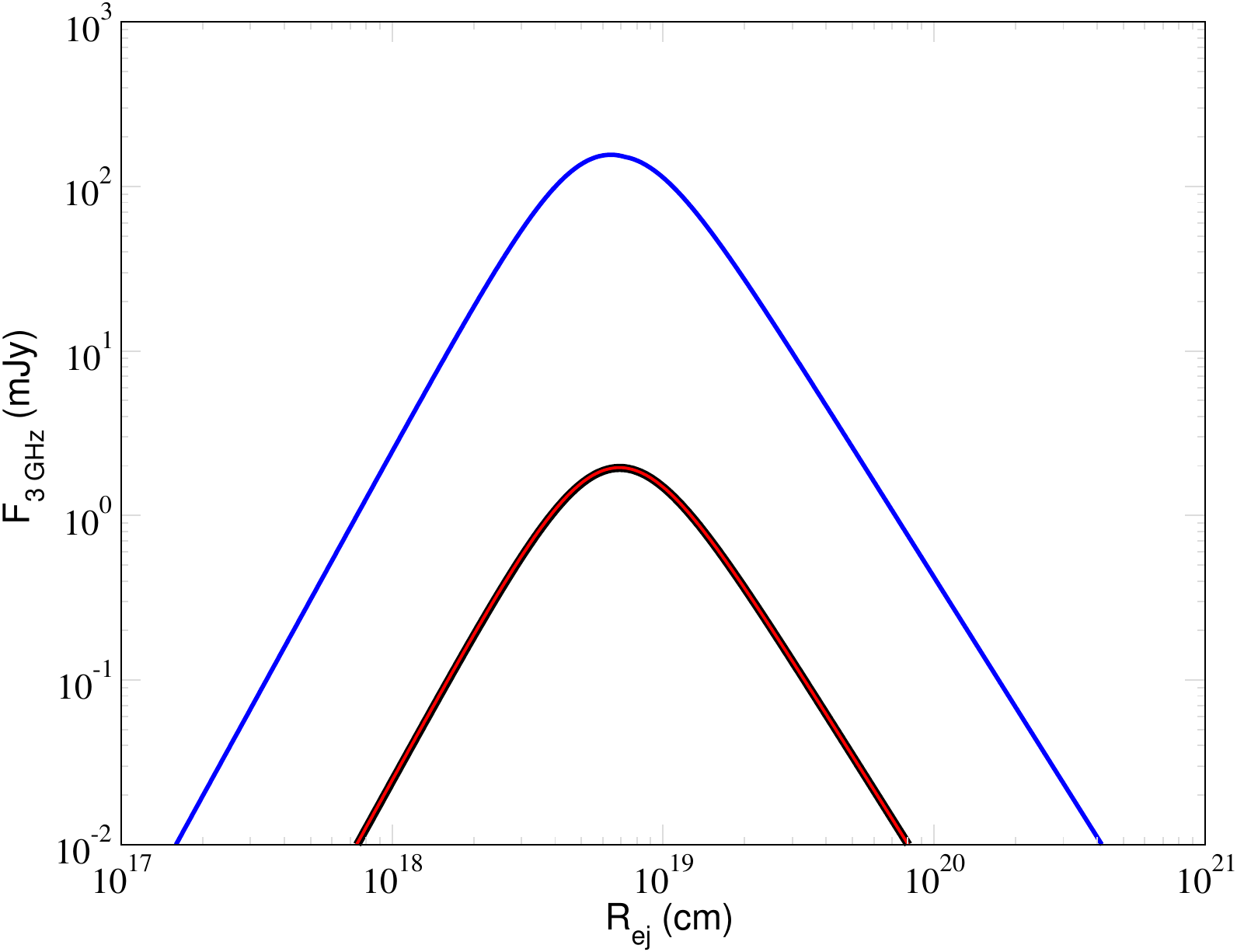}
\caption{Flux at 3 GHz as a function of age and ejecta size with $D=10$ Mpc for an SNR ($E_{51}=1$, $M_{\rm ej,\odot}$= 1, $n_0$= 1), a KNR ($E_{51}=1$, $M_{\rm ej,\odot}$= 0.1, $n_0$= 0.1), and a EKNR ($E_{51}=10$, $M_{\rm ej,\odot}$= 0.1, $n_0$= 0.1). We have chosen ${\bar{\epsilon}_{e,-1}}=1$ and $\epsilon_{B,-2}=1$. We have included the corrections by \citet{2020MNRAS.495.4981M} in the left panel of the figure to highlight the qualitative features of propagation through jet-evacuated material. We ignore this aspect in the rest of the calculation and the remaining figures. The red and black curves overlap in the right panel, which is a coincidence because the difference in typical density compensates for that in ejecta mass. }
\label{fig:F_3GHz}
\end{figure*}

The energetic electrons in the presence of magnetic field emit synchrotron emission \citep{Chevalier1982,Chevalier1998,2000ApJ...537..191F}. Assuming a fraction $\epsilon_B$ of shocked energy goes into amplifying the magnetic field, the comoving magnetic energy density in the non-relativistic regime is $\frac{B^2}{8\pi}=2m_{\rm p}\beta_{\rm ej}^2{\rm c}^2n_{\rm ISM}\epsilon_B$.
Substituting the expression for $\beta_{\rm ej}$ (Eq. \ref{eq:beta_ej}), the expression for the magnetic field becomes
\begin{equation}
\begin{aligned}
    B &=9.1\times 10^{-4}n_0^{1/2}\epsilon_{B,-2}^{1/2}E_{51}^{1/2}M_{\rm ej,\odot}^{-1/2} {\rm G},
    \hspace{0.5 cm} {\rm (coasting\hspace{0.1cm} phase)}\\
    B &=0.014\epsilon_{B,-2}^{1/2}E_{51}^{1/5}n_0^{3/10}t_{\rm yr}^{-3/5}{\rm G} \hspace{1.4cm} {(\rm ST\hspace{0.1cm}phase)},
 \end{aligned}   
 \label{eq:B_field}
\end{equation}
where $\epsilon_B\equiv 0.01\epsilon_{B,-2}$. The characteristic synchrotron frequency $\nu_m$ is given by $\nu_m=\gamma^2_m\nu_B$, where $\nu_B=\frac{{\rm e}B}{2\pi m_{\rm e}{\rm c}}=2.8\times 10^6 B$ Hz is the cyclotron frequency. For our fiducial cases of SNR and KNR, the solutions are in the deep Newtonian regime, while for EKNR, $\gamma_m\sim 10$ in the coasting phase (Eq. \ref{eq:gmtwocases}) and the deep Newtonian regime is typically obtained in the ST phase. In the deep Newtonian regime, the expression for $\nu_m$ is given by
\begin{equation}
\begin{aligned}
    \nu_m \sim 10^4n_0^{1/2}\epsilon_{B,-2}^{1/2}E_{51}^{1/2}M_{\rm ej,\odot}^{-1/2}\hspace{0.2cm} {\rm Hz},
    \hspace{1.5 cm} {\rm (coasting\hspace{0.1cm} phase)}\\
    \nu_m\sim 1.6\times 10^5\epsilon_{B,-2}^{1/2}E_{51}^{1/5}n_0^{3/10}t_{\rm yr}^{-3/5}\hspace{0.2cm}{\rm Hz} \hspace{1.5cm} {(\rm ST\hspace{0.1cm}phase)},
 \end{aligned}   
\end{equation}
for slow-cooling electrons, or $\nu_m<\nu_{\rm c,sync}$ \citep{SPN1998,GS2002}, where $\nu_{\rm c,sync}$ is the cooling frequency due to synchrotron losses alone (Appendix \ref{subsec:nu_c}). The power in the comoving frame due to one electron at the frequency $\nu_m$ is $P_{\nu_m}=3.8\times 10^{-22}\left(\frac{B}{\rm 1 G}\right) {\rm erg\hspace{0.1cm} s^{-1}Hz^{-1}}$. The number of shocked electrons (which is a Lorentz invariant) within a dynamical time is $\sim \frac{ 4\pi}{3} R^3\xi_e n_{\rm ISM}$. Therefore, we multiply this number with the power per electron to obtain the maximum optically thin synchrotron luminosity, $L_{\nu_{m,0}}\sim \xi_e n_{\rm ISM}\left(\frac{4\pi}{3}\right) R^3P_{\nu_{m,0}}$.
All together,
\begin{equation}
\label{eq:Lnum}
\begin{aligned}
    L_{\nu_{m,0}}\sim 9\times 10^{23}{\bar{\epsilon}_{e,-1}}\epsilon_{B,-2}^{1/2}E_{51}^{3}M_{\rm ej,\odot}^{-3}n_0^{3/2}t_{\rm yr}^{3}\hspace{0.2cm}\mbox{erg s}^{-1}\mbox{Hz}^{-1}\hspace{0.2cm} {(\rm coasting\hspace{0.1cm}phase),} \\
    L_{\nu_{m,0}}\sim 2\times 10^{32}{\bar{\epsilon}_{e,-1}}\epsilon_{B,-2}^{1/2}E_{51}^{6/5}n_0^{3/10}t_{\rm yr}^{-3/5}\hspace{0.5cm}  \mbox{erg s}^{-1}\mbox{Hz}^{-1}\hspace{0.6cm} {(\rm ST\hspace{0.1cm}phase).}
    \end{aligned}   
\end{equation}
The equation above applies in the deep Newtonian regime. Furthermore, for the physical situation being discussed, we verified that $\nu_m<\nu_{\rm sa}<\nu_{\rm c,sync}$, where $\nu_{\rm sa}$ is the synchrotron self-absorption frequency (Appendix \ref{subsec:nu_sa}). We caution that in this case, the luminosity at $\nu_m$ is not given by the expression given in Eq. \ref{eq:Lnum}. Nonetheless, as we are interested in luminosity or flux at $\nu>\nu_{\rm sa}>\nu_m$, we can still use these expressions to compute flux by extrapolation, as done below. We use the subscript `0' in the expressions for $L_{\nu_{m,0}}$ and $F_{\nu_{m,0}}$ to clarify this difference. The spectral flux at a given observation frequency $\nu$ with $\nu_m<\nu_{\rm sa}<\nu<\nu_{\rm c,sync}$ is given by $F_{\nu}=F_{\nu_{m,0}}\left(\frac{\nu}{\nu_m}\right)^{-(p-1)/2}$, which after substitution becomes
\begin{equation}
    \begin{aligned}
    F_{\nu}=5.4\times 10^{-7}\frac{{\bar{\epsilon}_{e,-1}}\epsilon_{B,-2}^{7/8}E_{51}^{27/8}M_{\rm ej,\odot}^{-27/8}n_0^{15/8}t_{\rm yr}^{3}}{\rm D_{10}^2}\left(\frac{\nu}{\rm 3\hspace{0.05cm} GHz}\right)^{-3/4}\hspace{0.2cm} {\rm mJy}\\ {(\rm coasting \hspace{0.1cm} phase)},\\
    F_{\nu}=8.85\times 10^2 \frac{{\bar{\epsilon}_{e,-1}}\epsilon_{B,-2}^{7/8}E_{51}^{27/20}n_0^{21/40}t_{\rm yr}^{-21/20}}{\rm D_{10}^2}\left(\frac{\nu}{\rm 3\hspace{0.05cm} GHz}\right)^{-3/4} \hspace{0.4cm} {\rm mJy}\\ \hspace{0.4cm} {(\rm ST \hspace{0.1cm} phase)},
    \end{aligned}
    \label{eq:flux_nu}
\end{equation}
where we have assumed Euclidean geometry, as appropriate for distances of $D\lesssim 10$ Mpc with $D_{10}=\frac{D}{\rm 10 Mpc}$. We normalised the expression for flux with respect to the central frequency of VLASS \citep{Gordon2023}. In Fig. \ref{fig:F_3GHz}, we plot the observed flux for the three classes of objects considered in this work. The peak of the observed flux occurs at $t_{\rm ST}$ with a peak flux of $\sim$ 1 mJy ($D=10$ Mpc) for both the SNR and KNR. EKNRs are significantly brighter and also have shorter $t_{\rm ST}$. This will significantly impact the detection prospects for such extreme objects; we discuss this topic in subsequent sections. We have also included the corrections by \cite{2020MNRAS.495.4981M} in this figure, which are briefly discussed in Sect. \ref{sec:dynamics_newtonian}. We note that the calculations discussed above are valid at $t>t_{\rm col}$, while at $t<t_{\rm col}$, there is an extremely sharp decrease in flux ($\sim t^{10}$) due to the decrease in external density caused by the previous propagation of the jet.  

\subsubsection{Flux at $t_{\rm ST}$}
As objects have the highest detection probability when they are brightest, it is useful to have the expression for the peak flux, which occurs at $t_{\rm ST}$ (Fig. \ref{fig:F_3GHz}). Using the expression for $t_{\rm ST}$ (Eq. \ref{eq:t_ST}) in the coasting phase solution for the observed flux (Eq. \ref{eq:flux_nu}), we have
\begin{equation}
   F_{\rm 3 GHz}^{\rm ST}=1.1\times 10^2\xi_e\frac{E_{51}^{7/8}M_{\rm ej,\odot}^{1/8}\left(\epsilon_{B,-2}n_0\right)^{7/8}}{D_{\rm 10}^2}\hspace{0.2cm} {\rm mJy},
\end{equation}
where we we keep $\xi_e$ explicit. Using the expression for $\xi_e$ in the deep Newtonian regime (Eq. \ref{eq:xie}), 
\begin{equation}
\label{eq:F@tST}
    F_{\rm 3 GHz}^{\rm ST}=2.4\frac{\bar{\epsilon}_{e,-1}E_{51}^{15/8}\left(\frac{\epsilon_{B,-2}n_0}{M_{\rm ej,\odot}}\right)^{7/8}}{D_{\rm 10}^2}\hspace{0.2cm} {\rm mJy} 
.\end{equation}
The peak flux is a strong function of ejecta energy, which explains the dramatic boost to observed flux in the case of a EKNR. The peak flux is a degenerate combination of $\epsilon_{B,-2}, n_0$, and $M_{\rm ej,\odot}$. This is visible in Fig. \ref{fig:F_3GHz} for the SNR and KNR cases, which both have $n_0/M_{\rm ej,\odot}=1$. 

\subsubsection{Peak flux at X-ray frequency}
We also compute the flux at $t_{\rm ST}$ in the X-ray band. 
The luminosity at 1 keV (or $\nu=2.4\times 10^{17}$ Hz) with $\nu>\nu_c$ is given by
$L_{\nu}=L_{\nu_{m,0}}\left(\frac{\nu_c}{\nu_m}\right)^{-(p-1)/2}\left(\frac{\nu}{\nu_c}\right)^{-p/2}
$. At $t=t_{\rm ST}$, 
\begin{equation}
     \nu F_{\nu}\vert_{\rm 1 keV}^{\rm ST}=0.62\times 10^{-13}\xi_e\frac{E_{51}^{5/8}\epsilon_{B,-2}^{1/8}n_0^{11/24} M_{\rm ej,\odot}^{1/24}}{D_{\rm 10 \hspace{0.05cm}Mpc}^2 (1+Y)}\hspace{0.2cm} \rm erg\hspace{0.1cm}cm^{-2}s^{-1}
.\end{equation}
Using the expression for $\xi_e$ in the deep Newtonian regime (Eq. \ref{eq:xie}), 
\begin{equation}
     \nu F_{\nu}\vert_{\rm 1 keV}^{\rm ST}=1.4\times 10^{-15}\frac{\bar{\epsilon}_{e,-1}E_{51}^{13/8}\epsilon_{B,-2}^{1/8}n_0^{11/24} M_{\rm ej,\odot}^{-23/24}}{D_{\rm 10 \hspace{0.05cm}Mpc}^2 (1+Y)}\hspace{0.2cm} \rm erg\hspace{0.1cm}cm^{-2}s^{-1}
.\end{equation}
In the equation above, and in the full results shown in Fig. \ref{fig:nu_Lnu}, we account for inverse-Compton losses affecting $\nu_c$ (Sect. \ref{subsec:ssc}). For our fiducial cases, $Y\lesssim 1$, and so synchrotron self-Compton losses are a minor correction. The flux limit for eROSITA in the soft X-ray band (0.5-2 keV) is $\sim 10^{-15}-10^{-14}$ ${\rm erg\hspace{0.1cm} cm^{-2}s^{-1}}$ \citep{erosita2012}. The ongoing Einstein probe \citep{Einstein2022} has a similar energy band with slightly worse flux limit, which depends on the length of exposure. Therefore, we expect to see objects in X-rays up to distances of a few megaparsecs, which can complement radio observations. If synchrotron is the dominant process in X-rays (as indeed expected; see Sect. \ref{subsec:ssc}), we expect a correlation between radio and X-ray flux, as both arise from the same population of electrons. 

\subsubsection{Synchrotron emission cutoff}
\label{subsec:sync_cutoff}
We compute the synchrotron cutoff by requiring that the diffusion length scale of electrons, while gyrating in the magnetic field, be less than the size of the remnant. The diffusion coefficient in the Bohm limit is given by $D={\rm c}r_g/3$,
where $r_g$ is the gyration radius, $r_g=\gamma m_{\rm e}{\rm c}^2/{\rm e}B,$
which can be simplified to $r_g=1.7\times 10^3\frac{\gamma}{B}$ cm. Imposing the condition as stated above, we have $Dt<\beta_{\rm ej}^2{\rm c}^2 t^2$,
or $t>\frac{D}{\beta_{\rm ej}^2 {\rm c}^2}$. We then use the synchrotron cooling timescale (Sect. \ref{subsec:nu_c}) in the expression to obtain
\begin{equation}
    \gamma_{\rm s,co}=2\times 10^8\beta_{\rm ej}B^{-1/2}.
\end{equation}
A more detailed calculation for the cutoff Lorentz factor is given by \cite{ZA2007}, which in the Bohm regime leads to a pre-factor of $3.6\times 10^7$ instead of $2\times 10^8$.
Using the value from  \cite{ZA2007}, the corresponding cutoff frequency is $\nu_{\rm s,co}=\gamma_{\rm s,co}^2\nu_B=3.6\times 10^{21}\beta_{\rm ej}^2\hspace{0.2cm}{\rm Hz}$. 
Plugging in $\beta_{\rm ej}$,
\begin{equation}
    \begin{aligned}
    h\nu_{\rm s,co}\sim 10^4\left(\frac{E_{\rm 51}}{M_{\rm ej,\odot}}\right)\hspace{0.1cm}{\rm eV}\hspace{1.3cm} ({\rm coasting\hspace{0.1cm} phase}), \\
    h\nu_{\rm s,co}\sim 4\times 10^6 E_{\rm 51}^{2/5}n_0^{-2/5}t_{\rm yr}^{-6/5}\hspace{0.1cm}{\rm eV}\hspace{0.8cm} ({\rm ST\hspace{0.05cm}phase}).
    \end{aligned}
\end{equation}
As $\beta_{\rm ej}$ is constant in the coasting phase, the corresponding expression gives a rough estimate at $t_{\rm ST}$ (at later times $\nu_{\rm s,co}$ only declines). For SNRs, the expected cutoff is of the order of $\sim 10$ keV, which becomes $\sim 100$ keV -- 1 MeV for our choice of KN parameters. Therefore, this feature, along with others (which we discuss later), can be used to separate SNRs from KNRs.

\subsection{Synchrotron self-Compton emission}
\label{subsec:ssc}
\begin{figure}
\centering 
\includegraphics[width=\columnwidth]{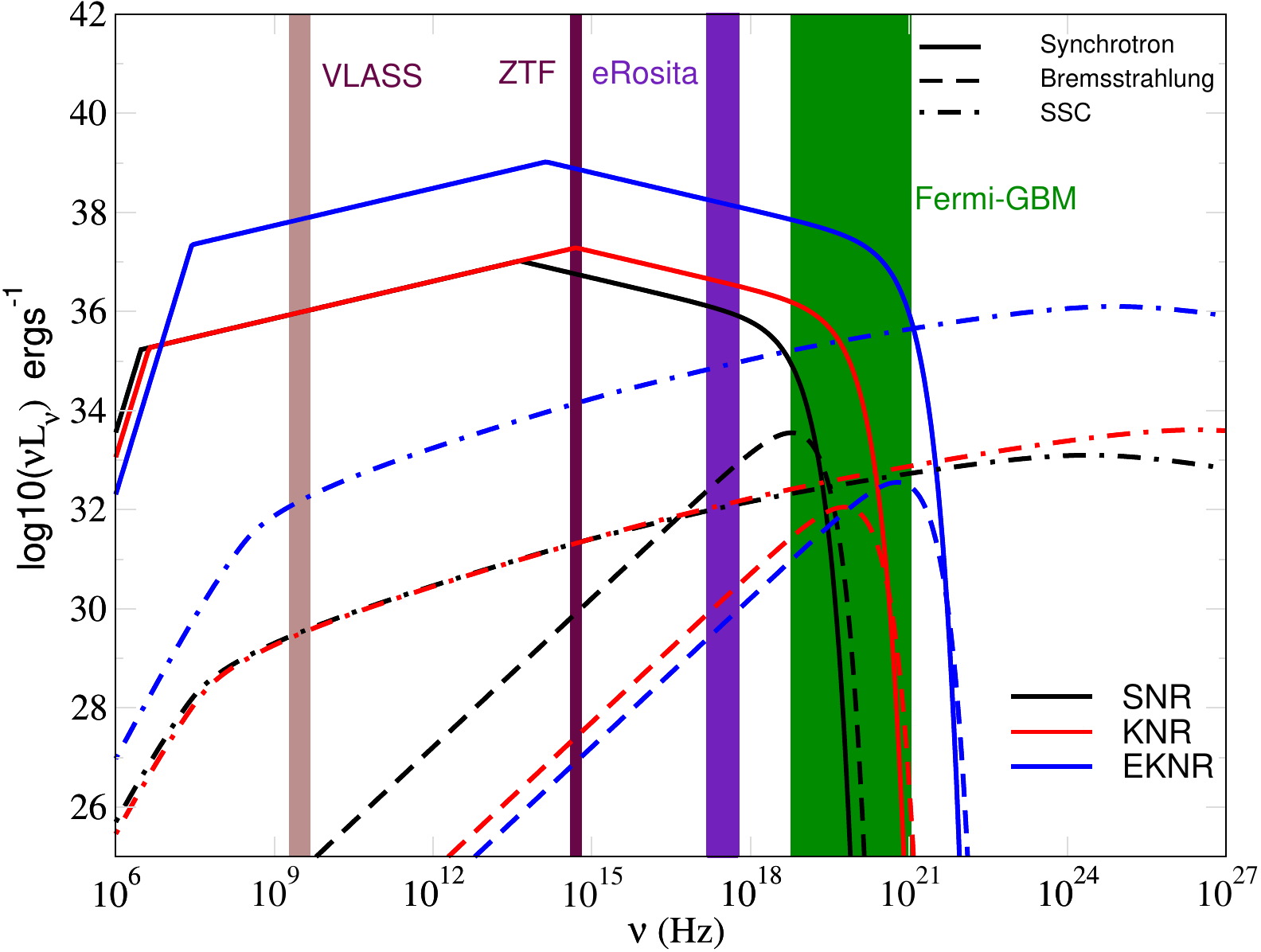}
\caption{ $\nu L_{\nu}$ as a function of $\nu$ at $t_{\rm ST}$ for a SNR (black), a KNR (red), and a EKNR (blue). }
\label{fig:nu_Lnu}
\end{figure}

In addition to synchrotron, electrons  also cool via Compton losses by the synchrotron photons, known as synchrotron self-Compton (SSC) loss. Therefore, Eq. \ref{eq:gamma_c_sync} is modified as
\begin{equation}
    \gamma_c m_{\rm e}{\rm c^2}=\frac{4}{3}\sigma_{\rm T}\gamma_c^2\left(\frac{B^2}{8\pi}+U_{\rm sync}\right){\rm c}t
    ,\end{equation}
where $U_{\rm sync}$ is the energy density of the synchrotron photon spectrum. As a result, the cooling Lorentz factor is modified as
  $  \gamma_c\rightarrow \frac{\gamma_{\rm c,sync}}{1+Y}$, where $Y$ is the Compton parameter whose expression is given by \cite{BNDP2015}, $ Y=\frac{\epsilon_e}{\epsilon_B(3-p)(1+Y)}\left(\frac{\nu_m}{\nu_c}\right)^{(p-2)/2}$, where $\nu_c$ is the cooling frequency with the Compton correction included, that is, $\nu_c=\nu_{\rm c,sync}/(1+Y)^2$,  where $\nu_{\rm c,sync}$ is the cooling frequency without SSC losses. Plugging in the expressions with $p=2.5$, we have 
\begin{equation}
Y(1+Y)^{0.5}=0.2\bar{\epsilon}_{e,-1}\epsilon_{B,-2}^{-3/4}E_{51}^{1/4}n_0^{1/3}M_{\rm ej,\odot}^{-1/12}
.\end{equation}
Using fiducial parameters, we obtain $Y(1+Y)^{0.5}\sim 0.18, 0.24$, and $0.43$ for SNRs, KNRs, and EKNRs, respectively. 
Therefore, Compton cooling has a minor effect on the scenarios that we consider. However, if $\epsilon_{B,-2}$ is much smaller than 1, then this effect is moderately important. Nevertheless, we include this effect in our calculation of the synchrotron spectrum as shown in Fig. \ref{fig:nu_Lnu}. The Compton-scattered frequency corresponding to $\nu_c$ is given by $\nu_{ c,{\rm  IC}}\sim \gamma_c^2\nu_c\sim \gamma_c^4\nu_B$. The value of $\gamma_{c}$ at $t_{\rm ST}$ is of the order of $10^5$.
Therefore,  $\nu_{ c,\rm {IC}}\sim 10^{24}$ Hz \footnote{This is slightly below the Klein-Nishina  cutoff in Fig. \ref{fig:nu_Lnu}, but not by a large margin. }. We solve for the SSC flux as a function of frequency, as explained in Appendix \ref{subsec:ssc_flux}.

In Fig. \ref{fig:nu_Lnu}, we plot $\nu L_{\nu}$ for synchrotron, SSC, and bremsstrahlung at $t_{\rm ST}$ for our fiducial cases. We find that bremsstrahlung (Appendix \ref{subsec:bremsstrahlung}) is subdominant compared to the other channels over the entire frequency domain. For SNRs, with larger $n_0$, bremsstrahlung may dominate over synchrotron in the X-ray band because it is highly sensitive to the value of the gas density. For our fiducial parameters of KNR, the synchrotron emission is further boosted compared to bremsstrahlung. Therefore, we  expect the X-ray flux to be dominated by synchrotron, which may be a diagnostic tool to separate it from SNRs, especially with the synchrotron cut-off feature (Sect. \ref{subsec:sync_cutoff}). The SSC process completely dominates the emission in the gamma-ray band. For KNRs, the overall transition from synchrotron to SSC happens around 1-10 MeV. In this context, we briefly discuss $\gamma$-ray detectability with  the Fermi gamma-ray burst monitor (Fermi-GBM). The sensitivity of this instrument is of the order of $0.5{\rm cm^{-2}s^{-1}}$ between 10 keV and 25 MeV. We note that this sensitivity is given in terms of photon flux, which is the number of photons per unit area per second. For the EKNR case in Fig. \ref{fig:nu_Lnu}, the flux will be of the order of $\sim 10^{-7}{\rm cm^{-2}s^{-1}}$ at 100 keV for an object located at $D=10$ Mpc. Therefore, it would be difficult to detect features in the gamma-ray part of the spectrum. 
We also show the relevant energy range of ZTF in Fig. \ref{fig:nu_Lnu}. Using its magnitude limit, we find that we can see objects up to a few megaparsecs. In subsequent sections, we concentrate on the radio part of spectrum and discuss how the light curves (Sect. \ref{subsec:lightcurve}) may help us distinguish KNRs from SNRs.   

\section{Number of detectable remnants}
\label{sec:N_remnants}
We now consider the detection prospects of KNRs with ongoing and future telescope surveys.
The very large area sky survey (VLASS) is observing the entire sky north of $-40^{\rm o}$ (33885 ${\rm deg^2}$)  in the ${\rm 2 GHz}\lesssim \nu \lesssim {\rm 4 GHz}$  band with an angular resolution of the order of 3 arcsec. The current sensitivity is of the order of 1 mJy \citep{Gordon2021}, which will eventually reach the level of 0.1 mJy. VLASS will cover its footprint over three epochs separated by 32 months. For the current flux threshold of $\sim 1$ mJy, the expected distance up to which a KN can be seen is $\sim 10$ Mpc (Fig. \ref{fig:F_3GHz}). For a generic flux threshold $F_{\rm ths}$, the maximum distance is given by $D^{\rm max}=\left(L_{\rm 3 GHz}/4\pi F_{\rm ths}\right)^{1/2}$,
which can be rewritten as
   $ D_{10}^{\rm max}=\left(\frac{F_{\rm 3 GHz,10}}{F_{\rm ths}}\right)^{1/2}$,
where $F_{\rm 3 GHz,10}$ is the flux at 10 Mpc.
Objects of a given age $t$ are detected at a rate of
\begin{equation}
\label{eq:dNdt}
    \frac{{\rm d}N_{\rm obs}}{{\rm dt}}={\mathcal {R}(t)}\frac{4\pi}{3}D^3,
\end{equation}
where ${\mathcal {R}(t)}$ is the rate of formation of KNe per volume per time, which is $\sim 300$ Gpc$^{-3}$yr$^{-1}$ \citep{2023PhRvX..13a1048A}. Substituting the numbers, we have, $\frac{{\rm d}N_{\rm obs}}{{\rm dt}}=1.2\times 10^{-3}(D_{\rm 10}^{\rm max})^3\hspace{0.2cm} {\rm yr^{-1}}$.
For $L_{\nu}\propto t^{\alpha}$, $\frac{{\rm d}N_{\rm obs}}{{\rm dt}}\propto  t^{3\alpha/2}$
or equivalently $ \frac{{\rm d}N_{\rm obs}}{{\rm dlogt}}\propto t^{1+3\alpha/2}$. While Eq. \ref{eq:dNdt} is denoted in terms of the observer's time, the luminosity is a function of the age of the system. In practice, we have many systems with different ages and we are averaging over their properties at a given observer time. Under our assumptions, this does not change our results.
For the coasting phase regime, $\alpha=3,$ while for the ST regime, $\alpha=-21/20$ (Eq. \ref{eq:flux_nu}). Substituting, we have 
\begin{equation}
\begin{aligned}
\label{eq:Nobsvst}
  \frac{{\rm d}N_{\rm obs}}{{\rm dlogt}}\propto t^{11/2} \hspace{0.4cm} (\rm coasting\hspace{0.1cm} phase), \\
  \frac{{\rm d}N_{\rm obs}}{{\rm dlogt}}\propto t^{-1/2} \hspace{0.8cm} (\rm ST\hspace{0.2cm} phase).
   \end{aligned}
\end{equation}
For the fiducial KNR parameters, $D_{10}^{\rm max}=1.5$.

\begin{figure}
\centering 
\begin{subfigure}[b]{0.5\textwidth}
\includegraphics[scale=0.3]{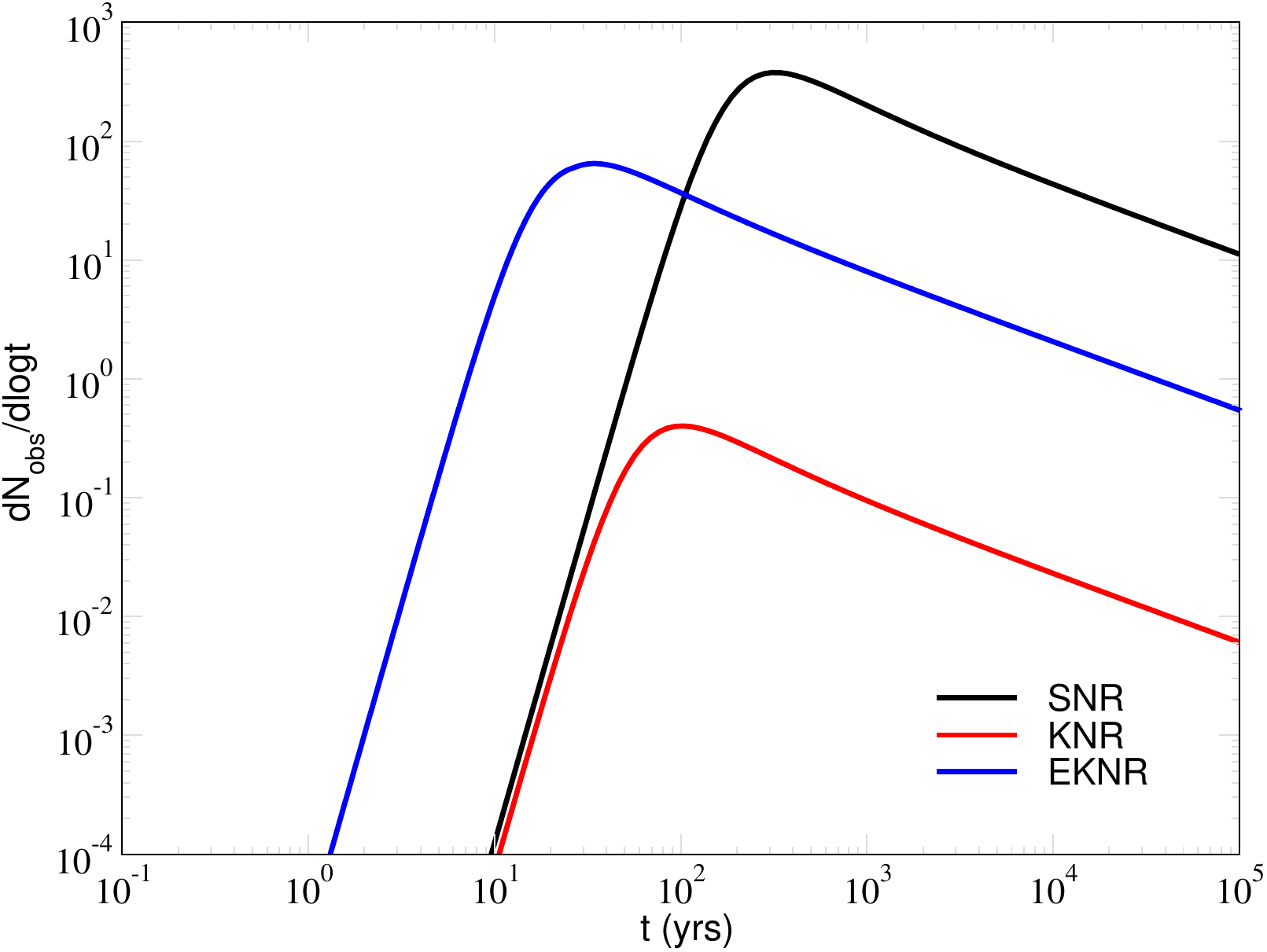}
\end{subfigure}
\begin{subfigure}[b]{0.5\textwidth}
\includegraphics[scale=0.3]{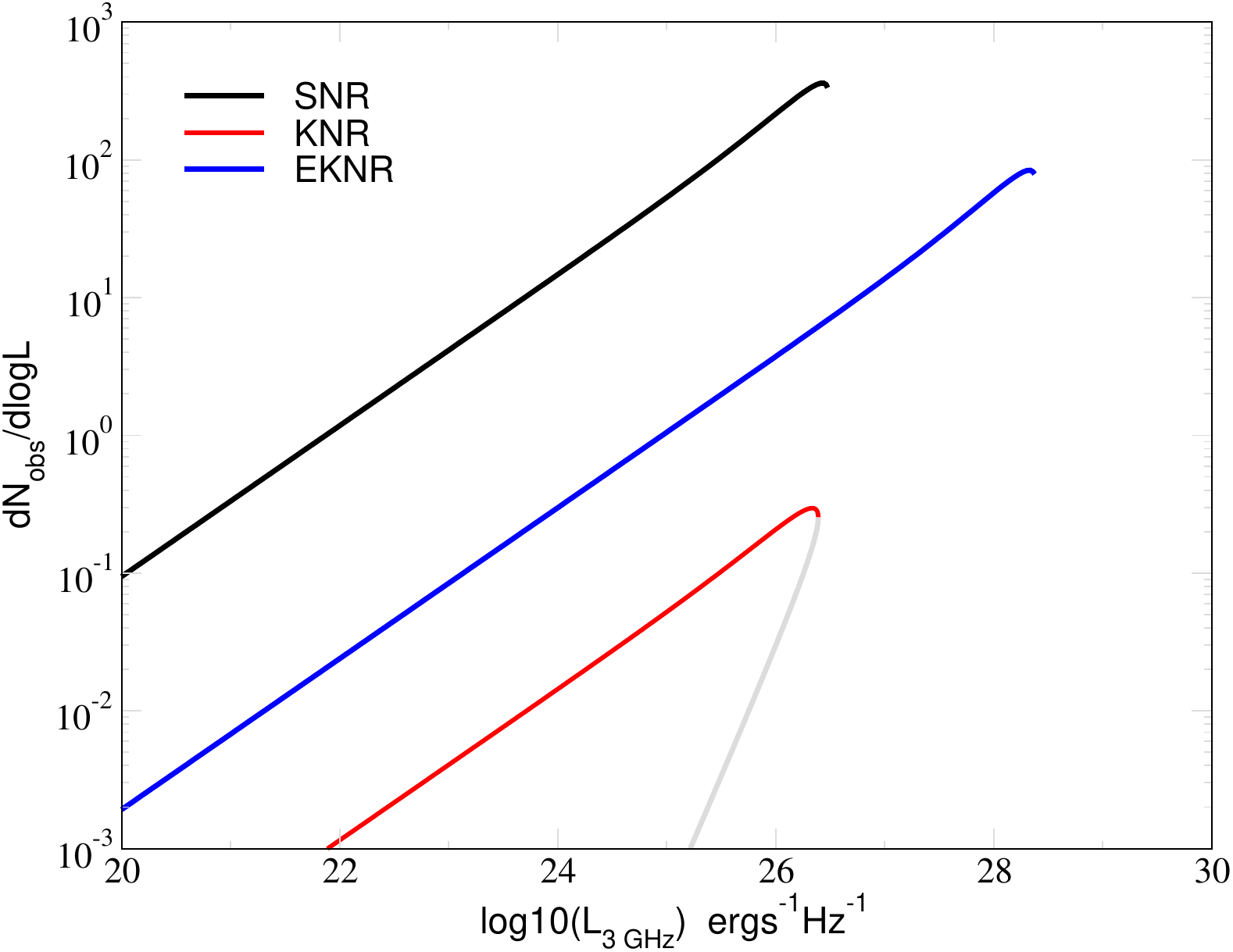}
\end{subfigure}
\caption{Abundance of SNR, KNR and EKNRs. Top: Number of systems observed by an all-sky radio survey per logarithmic unit of age, $\frac{{\rm d}N_{\rm obs}}{{\rm dlog}t}$, for SNR, KNR, and EKNRs. Bottom: $\frac{{\rm d}N_{\rm obs}}{{\rm dlog}L}$ for the same cases as in the upper panel. For the SN case, we show $\frac{{\rm d}N_{\rm obs}}{{\rm dlog}L}$ for the coasting regime in grey in order to highlight its non-monotonic nature. We assume a minimum flux threshold of $F_{\rm ths}=1$ mJy, an observation frequency of 3 GHz, and fiducial ejecta parameters as in Fig. \ref{fig:F_3GHz}. EKNRs evolve on a short-enough timescale for their evolution to be observable (as explored in more detail in Fig. \ref{fig:F_curvature_1}). }
\label{fig:dN_dt}
\end{figure}

We expect the quantity $\frac{{\rm d}N_{\rm obs}}{{\rm dlogt}}$ to be dominant at the Sedov time. At $t_{\rm ST}$, we can use Eq. \ref{eq:F@tST} to obtain the general expression,
\begin{align}
    \frac{{\rm d}N_{\rm obs}}{{\rm dt}}=0.004\left(\frac{{\mathcal {R}(t)}}{\rm 300 Gpc^{-3}yr^{-1}}\right)\bar{\epsilon}_{e,-1}^{\frac{3}{2}}E_{51}^{\frac{45}{16}}\left(\frac{\epsilon_{B,-2}n_0}{M_{\rm ej,\odot}}\right)^{\frac{21}{16}}
    \left(\frac{F_{\rm ths}}{\rm 1 mJy}\right)^{-\frac{3}{2}}\hspace{-0.2cm}{\rm yr^{-1}}.
    \label{eq:dN_dt_full}
\end{align}
For our fiducial KNR case, $t_{\rm ST}\sim 50$ yr or $\frac{{\rm d}N_{\rm obs}}{{\rm d\log t}}\sim 0.2,$ meaning that there is a roughly 20\%\ chance that such an object exists within the current VLASS dataset. In the future, with a reduction of the VLASS flux threshold to 0.1 mJy \citep{Gordon2023}, these numbers will be boosted by a factor of about $30$. In Fig. \ref{fig:dN_dt}, we plot $\frac{{\rm d}N}{{\rm dlog}t}$ for SNRs, KNRs, and EKNRs with our choice of fiducial parameters. We also consider a few cases with a distribution over density and ejecta mass of EKNRs, which we describe below. We take the SN occurrence rate to be $\sim 10^5\mbox{ Gpc}^{-3}\mbox{ yr}^{-1}$ \citep{2011MNRAS.412.1441L}, which is approximately 300 times higher than the KN rate. A consistency check for the number of detectable SNRs given our assumptions is provided in Appendix \ref{sec:R_SN}.

While Eq. \ref{eq:Nobsvst} is written in terms of time, for an object at a known distance, the luminosity is directly inferred from observations. Therefore, it is useful to convert the expression into one in terms of luminosity using the relation, $\frac{{\rm d}N_{\rm obs}}{{\rm d}L}=\frac{{\rm d}N_{\rm obs}}{{\rm d}t}\left(\frac{{\rm d}L}{{\rm d}t}\right)^{-1}$. Using the dependence of luminosity on time, we have
    $L\left(\frac{{\rm d}N_{\rm obs}}{{\rm d}L}\right)=\frac{{\rm d}N_{\rm obs}}{{\rm d}t}\frac{t}{\alpha}$. In terms of luminosity, 
\begin{equation}
\begin{aligned}
  \frac{{\rm d}N_{\rm obs}}{{\rm dlogL}}\propto L^{11/6} \hspace{0.2cm} (\rm coasting\hspace{0.1cm}phase), \\
  \frac{{\rm d}N_{\rm obs}}{{\rm dlogL}}\propto L^{1/2} \hspace{0.8cm} (\rm ST\hspace{0.2cm} phase).
   \end{aligned}
\end{equation}
In the bottom panel of Fig. \ref{fig:dN_dt}, we plot the same as in the upper panel but in terms of luminosity. We only consider the ST regime for this plot because $\frac{{\rm d}N}{{\rm dlog}L}$ is a non-monotonic function of $L$ and the number of objects in the ST regime dominates over the number of sources in the coasting phase (see the thin grey line in the figure).

If we are able to spatially resolve the remnants, we can directly track their evolution in size as a function of time (see e.g. \citealt{2016MNRAS.462L..31B}). This provides additional information and can inform us as to whether the object is in the coasting or ST phase. In Fig. \ref{fig:F_3GHz} (right panel), we plot the observed flux as a function of the size of the ejecta. We find that SNRs and KNRs are degenerate, with the Sedov radii of all three cases being similar. Nevertheless, observation of the ejecta evolution over time can help distinguish remnants from potential contaminants. The related observable is the surface brightness $S_{\nu}$ given by $S_{\nu}=\frac{L_{\nu}}{4\pi R_{\rm ej}^2}$. The observed distribution of $S_{\nu}$ is given by
\begin{equation}
    \frac{{\rm d}S_{\nu}}{{\rm dt}}={\mathcal {R}(t)}\frac{4\pi}{3}\left(\frac{L_{\nu}}{4\pi F_{\rm ths}}\right)^{3/2}\left(\frac{L_{\nu}}{4\pi R_{\rm ej}^2}\right).
\end{equation}
As the size of a KNR varies as $t$ and $t^{2/5}$ in blast-wave and ST-regime, respectively,
\begin{equation}
    \begin{aligned}
  \frac{{\rm d}S_{\nu}}{{\rm dlogt}}\propto t^{13/2} \hspace{0.2cm} (\rm coasting\hspace{0.1cm}phase), \\
  \frac{{\rm d}S_{\nu}}{{\rm dlogt}}\propto t^{-23/10} \hspace{0.6cm} (\rm ST\hspace{0.2cm} phase).
   \end{aligned}
\end{equation}
Given the flux sensitivity of a survey, we can see objects up to the maximum distance $D_{10}^{\rm max}$, as introduced above. The physical size of a resolvable object is given by $R_{\rm rsl}=4.5\times 10^{20}\left(\frac{\theta_{\rm rsl}}{3''}\right)D_{10}^{\rm max}\hspace{0.2cm} {\rm cm,}$ where $\theta_{\rm rsl}$ is the angular resolution of the survey, which we have normalised to 3 arcsec, for applicability to VLASS. As the Sedov radius of SNRs and KNRs are of the order of $10^{19}$ cm, in order to resolve a KNR, we have the relation
\begin{equation}
   \left(\frac{\theta_{\rm rsl}}{3''}\right)D_{10}^{\rm max}\lesssim 0.02.
   \label{eq:resolution_limit}
 \end{equation}
We can rewrite this equation in terms of $F_{\rm ths}$ as
\begin{equation}
    \left(\frac{\theta_{\rm rsl}}{3''}\right)F_{\rm ths}^{-1/2}\lesssim 0.02F_{\rm 3GHz,10}^{-1/2}
,\end{equation}
where we can use Eq. \ref{eq:F@tST} for the expression of $F_{\rm 3GHz,10}$ at $t_{\rm ST}$. Assuming a conservative value of $\theta_{\rm rsl}=1''$, we can resolve KNRs up to $\sim 600$ kpc (Eq. \ref{eq:resolution_limit}). We find the number of objects within 600 kpc to be $\frac{{\rm d}N_{\rm obs}}{{\rm dlogt}}\sim 10^{-5}$ at $t_{\rm ST}$, even when assuming ${\mathcal {R}(t)}$ to be ten times higher than the assumed value in Eq. \ref{eq:dNdt}. Therefore, we do not expect to find a spatially resolvable KNR with a survey like VLASS. 

\begin{figure}
\centering 
\begin{subfigure}[b]{0.5\textwidth}
\includegraphics[scale=0.3]{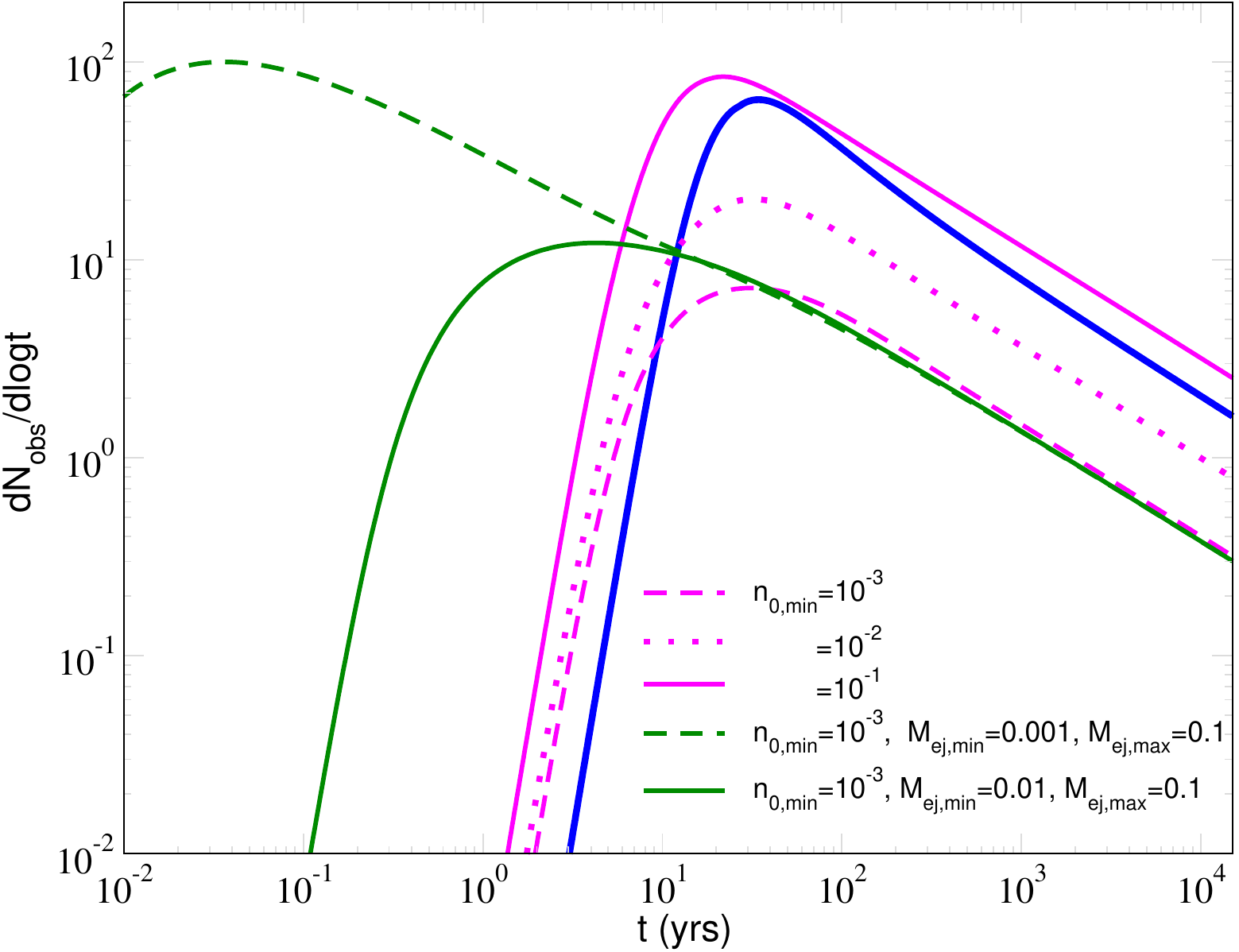}
\end{subfigure}
\begin{subfigure}[b]{0.5\textwidth}
\includegraphics[scale=0.3]{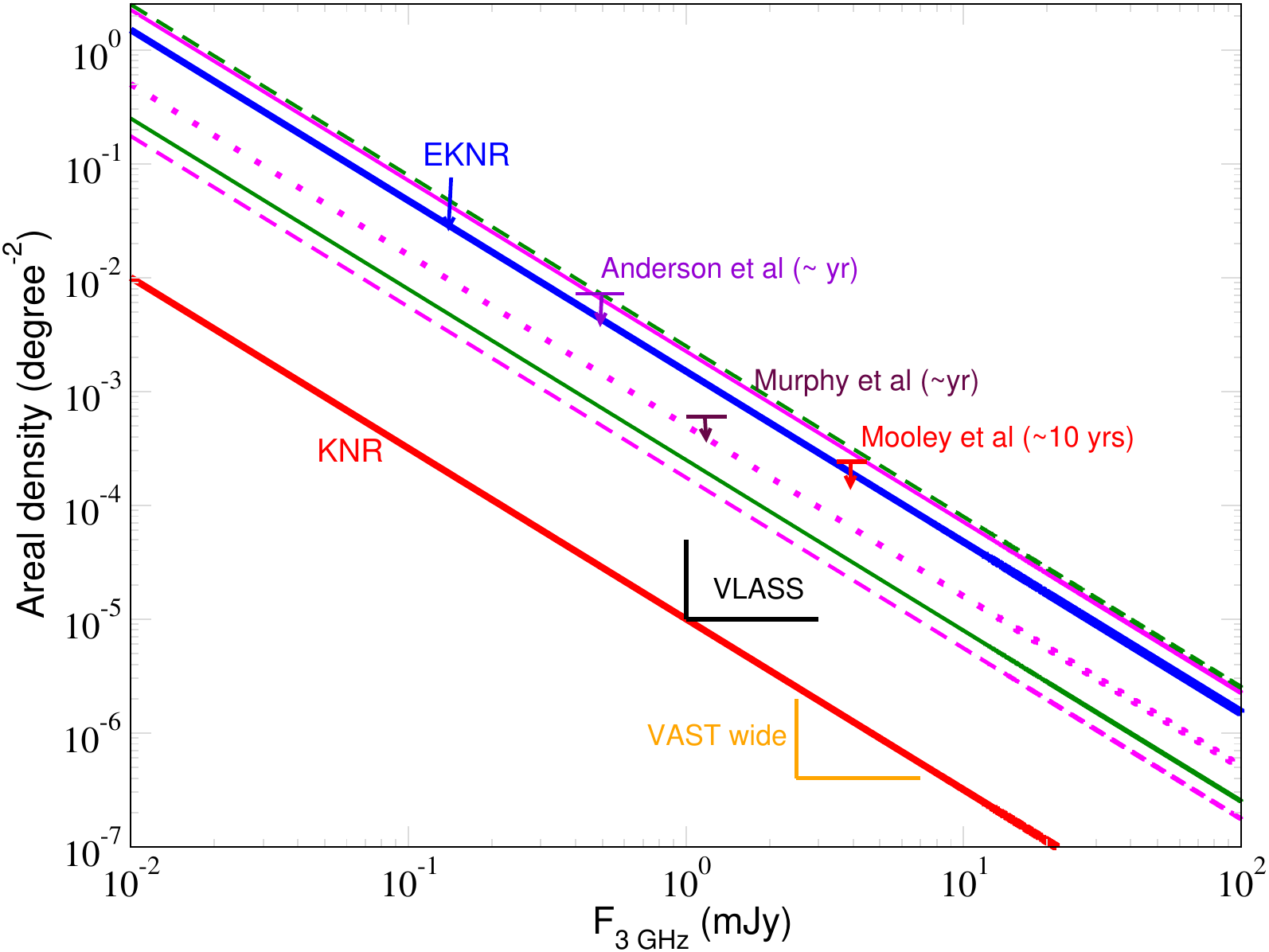}
\end{subfigure}
\caption{Dependence of EKNR abundance on surrounding density and mass distribution. Top:$\frac{{\rm d}N_{\rm obs}}{{\rm dlog}t}$ for EKNRs taking a distribution of external densities $n_0$ with $n_{\rm 0,max}=1\mbox{ cm}^{-3}$ and varying $n_{\rm 0,min}$ and $a$ (Eq. \ref{eq:dP_dn_dist}) shown by different line styles. In green, we consider a distribution over both density and ejecta mass but not over ejecta energy. For reference, we show our fiducial case of EKNR as in Fig. \ref{fig:dN_dt} in solid blue lines. Bottom: Areal density for the same cases as in the top panel. We also show the KNR case from Fig. \ref{fig:dN_dt}, which is a factor of about $ 10^2-10^3$ smaller than the EKNR cases. We have used the results of \citealt{Anderson2020}, \citealt{Murphy2021}, \citealt{Mooley2022} as upper limits, because these authors do not detect a KNR.  }
\label{fig:dN_dt_distribution}
\end{figure}

\subsection{Extended distribution over density and ejecta mass}
\label{sec:extended_dist}
Up to this point, we have considered single values of the external density and ejecta mass for KNe. Here, we consider a distribution of external density, and we discuss the same for ejecta mass below. We assume the external density of EKNRs to be
\begin{equation}
    \frac{{\rm d}P}{{\rm d}n_0}=An_0^{-a}.
    \label{eq:dP_dn_dist}
\end{equation}
The prefactor $A$ is fixed by the normalisation 
$\int_{\rm n_{0,min}}^{\rm n_{0,max}}\frac{{\rm d}P}{{\rm d}n_0}{\rm d}n_0=1$ or $A=\left(\int_{\rm n_{0,min}}^{\rm n_{0,max}}n_0^{-a}dn_0\right)^{-1}$. The rate of detectable objects is
\begin{equation}
    \frac{{\rm d}N}{{\rm d}t}={\rm R}(t)\frac{4\pi}{3}\int_{\rm n_{0,min}}^{\rm n_{0,max}}D(n_0)^3\frac{{\rm d}P}{{\rm d}n_0}{\rm d}n_0,
\end{equation}
where $D$ is a function of other ejecta parameters as well as the flux threshold, but we have suppressed them for brevity. We compute this quantity numerically to obtain the results shown in Fig. \ref{fig:dN_dt_distribution}. Here, we give a simple estimate of how $\frac{{\rm d}N}{{\rm d}t}$ changes at $t_{\rm ST}$ depending on the value of $a$. As $L_{\nu}\propto$ $n_0^{7/8}$ at $t_{\rm ST}$ (Eq. \ref{eq:F@tST}) and the observable  distance $D^{\rm max}\propto L_{\nu}^{1/2}$, the rate of detectable objects is
\begin{equation}
    \frac{{\rm d}N}{{\rm d}t}\sim \int n_0^{\frac{21}{16}-a}dn_0\sim n_0^{\frac{37}{16}-a}
.\end{equation}
As $t_{\rm ST}\sim n_0^{-1/3}$, the number of detected objects at $t_{\rm ST}$ is approximately $N\sim n_0^{95/48-a}$. We do not know the exact value of $a$ at present. To motivate a guess for $a$, we assume $n_{0,\rm min}=10^{-3}, n_{\rm 0,max}=1$ cm$^{-3}$ and require that the number of objects at $n=1$ cm$^{-3}$ is of the order of 1 percent of the population at $n_0=10^{-3}$ (this is  guided by observations of short GRB afterglows by \citealt{CBK2020}, as well as by the inferred gravitational wave delay time distribution between formation and merger of binary neutron star mergers; see \citealt{2024ApJ...966...17B}). This rough estimate gives $a=5/3$. Incorporating this value into Eq. 32, the number of detectable objects goes as $N\propto n_0^{15/48}$. Therefore, a small number of objects with large external density can naturally dominate the number of detectable objects. The total number of detected objects depends on $n_{\rm 0,min}$, ${\rm n_{0,max}}$, and $a$ through the normalisation parameter $A$. We consider a few cases where we fix $n_{\rm 0,max}=1$ and vary $a$ and $ n_{\rm 0,min}$ such that the number of object at $n=1$ cm$^{-3}$ is of the order of 1 percent of the population at $n_{\rm 0,min}$. We plot these cases in Fig. \ref{fig:dN_dt_distribution}. 
We see that the number of detected objects depends sensitively on our choice of the density distribution.

We also consider a distribution of ejecta masses with, 
\begin{equation}
    \frac{{\rm d}P}{{\rm d}M_{\rm ej}}=BM_{\rm ej}^{-b},
\end{equation}
where $B$ is the normalisation parameter. We take the lower and upper limit of ejecta mass to be 0.01 and 0.1 $M_{\odot}$, respectively. Without proper knowledge of the value of $b$, we choose $b=1,$ which corresponds to equal logarithmic contributions. Similar steps as above lead to 
\begin{equation}
    \frac{{\rm d}N}{{\rm d}t}={\rm R}(t)\frac{4\pi}{3}\int_{\rm n_{0,min}}^{\rm n_{0,max}} \int_{\rm M_{\rm ej,min}}^{\rm M_{\rm ej,max}}D(n_0,M_{\rm ej})^3\frac{{\rm d}P}{{\rm d}n_0}\frac{{\rm d}P}{{\rm d}M_{\rm ej}}{\rm d}n_0{\rm d}M_{\rm ej},
\end{equation}
We solve this expression numerically and plot it in Fig. \ref{fig:dN_dt_distribution}. We note that for extreme KNe with $M_{\rm ej,\odot}=0.01$, the bulk velocity is mildly relativistic and we use the dynamical expressions provided in Appendix \ref{subsec:dynamics_relativistic}. Still, some of the aspects can be understood from scaling relations obtained in the non-relativistic regime. While the flux at $t_{\rm ST}$ increases with decreasing $M_{\rm ej}$ (Eq. \ref{eq:F@tST}), the Sedov time itself is also a strong function of $M_{\rm ej}$ (Eq. \ref{eq:t_ST}). Therefore, the two effects compensate for each other in the quantity $\frac{{\rm d}N_{\rm obs}}{{\rm dlog}t}$ and the ejecta mass does not have a huge effect on the number of detectable objects at $t_{\rm ST}$. Reducing $M_{\rm ej}$ further has a much more drastic effect, with the ejecta moving at relativistic speeds, and causes a huge increase in the expected number of detected objects with extremely short $t_{\rm ST}$.

From the above calculations, we find that we should be in a position to detect a KNR in the near future with a VLASS-like survey with an achievable goal of $F_{\rm ths}\sim 0.1$ mJy. Even more interestingly, if EKNRs constitute a significant population of all KNRs, we expect to find 10-100 such objects. If we do not find any in the survey, we should be able to put strong constraints on the population of EKNRs. As no such candidate objects are known in the literature to our knowledge, we take the conservative approach and obtain an upper limit on the abundance of EKNRs (Sect. \ref{sec:ConstrainsnoEKN}). We discuss the potential signatures that can be used to   differentiate these objects from SNRs in Sect. \ref{sec:detection_prospects}. 

We also plot the areal density, or the number of objects per square degree, for the objects considered in Fig. \ref{fig:dN_dt_distribution}. For the number of objects, we consider the peak of the distribution of $\frac{{\rm d}N_{\rm obs}}{\rm dlogt}$. The number of objects is proportional to $(D^{\rm max})^3$, which is turn is proportional to $F^{-3/2}$ for an object of given luminosity. This gives rise to the qualitative behaviour seen in the bottom panel of Fig. \ref{fig:dN_dt_distribution}.

\subsection{Constraints on extreme KN abundance}
\label{sec:ConstrainsnoEKN}
In order to constrain the abundance of EKNRs, we show presently available constraints from transient surveys in the bottom panel of Fig. \ref{fig:dN_dt_distribution}. These are dedicated surveys that do not report a single KNR. Therefore, we use their results to put upper limits on the abundance of EKNRs. Presently, the strongest constraints come from the observations of \cite{Murphy2021}, but they strongly depend on the assumed density and mass distribution of EKNRs, as can be seen in Fig. \ref{fig:dN_dt_distribution}.
For our fiducial case ($E_{51}=10$, $M_{\rm ej,\odot}$= 0.1, $n_0$= 0.1), we find $f_{\rm EKN}\lesssim 0.3$, where $f_{\rm EKN}$ is the fraction of EKNRs with respect to the overall KNR population. As this constraint essentially depends on the value of $t\frac{\rm dN_{obs}}{\rm dt}$ at $t_{\rm ST}$, we derive a general expression using Eq. \ref{eq:dN_dt_full} and \ref{eq:t_ST}:
\begin{equation}
    f_{\rm EKN}\lesssim 0.3\left(\frac{{\mathcal {R}(t)}}{\rm 300 Gpc^{-3}yr^{-1}}\right)^{-1}\bar{\epsilon}_{e,-1}^{-\frac{3}{2}}\epsilon_{B,-2}^{-\frac{21}{16}}E_{52}^{-\frac{37}{16}}n_{-1}^{-\frac{47}{48}}M_{\rm ej,\odot,-1}^{\frac{23}{48}}\left(\frac{F_{\rm ths}}{\rm 1 mJy}\right)^{\frac{3}{2}}.
\end{equation}

\subsection{Future constraints}
In Fig. \ref{fig:dN_dt_distribution}, we also plot future constraints expected from VAST wide \citep{Murphy2021} and VLASS \citep{Mooley2016,Gordon2021}. The expected areal densities from these surveys are $4\times 10^{-7}$ at 2.5 mJy (1.4 GHz) and $10^{-5}$ at 1 mJy (3 GHz), respectively. These surveys are expected to find EKNRs if they constitute a significant fraction of the total KNR population. If no candidate objects are found in these surveys, we should be able to put strong constraints on the occurrence rate  of EKNRs  irrespective of their assumed density and mass distribution.

\section{Detection prospect of kilonova remnants}
\label{sec:detection_prospects}

\subsection{Estimating $t_{\rm ST}$ from the observed light curve}
\label{subsec:lightcurve}
\begin{figure}
\centering
\begin{subfigure}[b]{0.5\textwidth}
\includegraphics[scale=0.3]{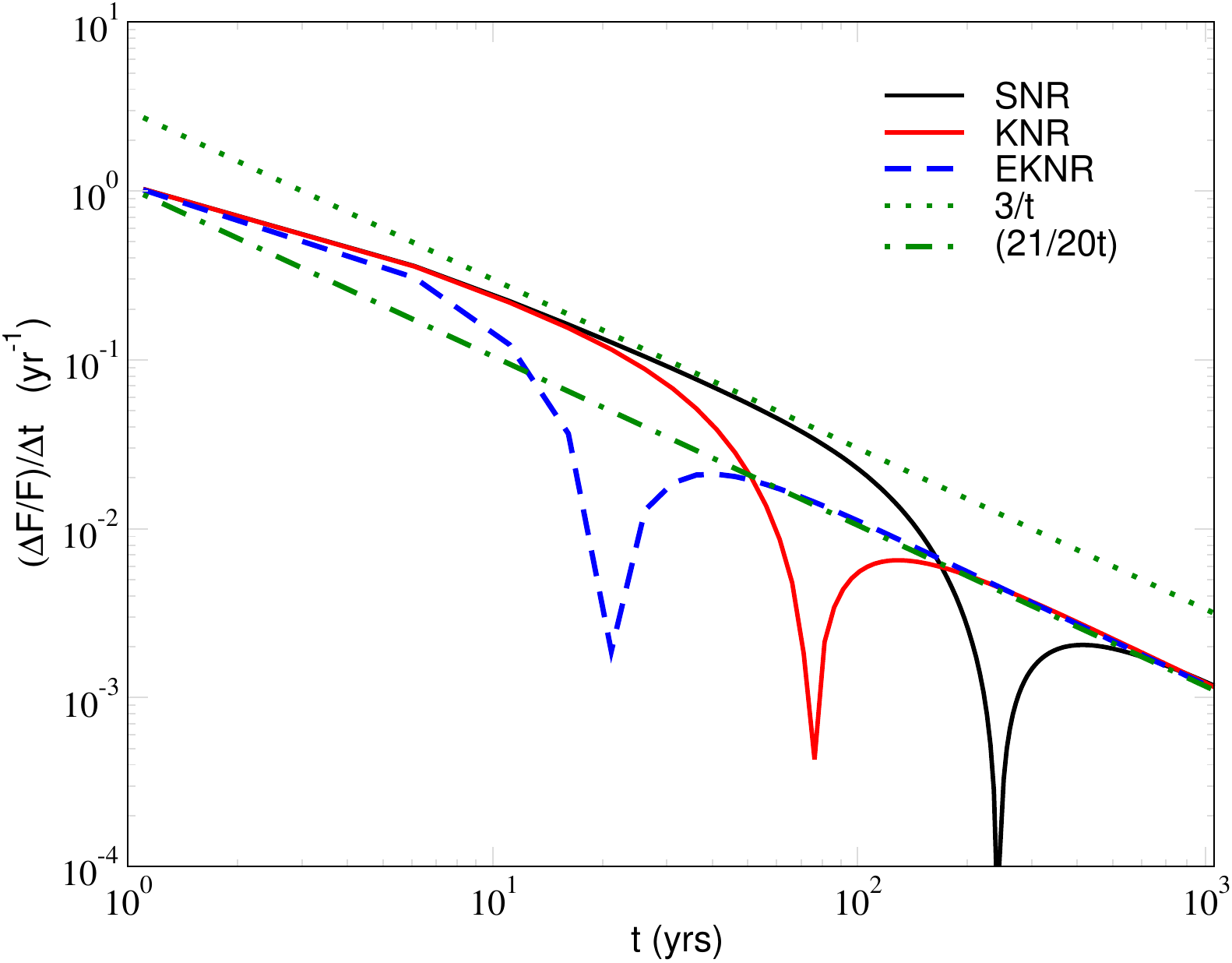}
\label{fig:F_curvature}
\end{subfigure}
\begin{subfigure}[b]{0.5\textwidth}
\includegraphics[scale=0.3]{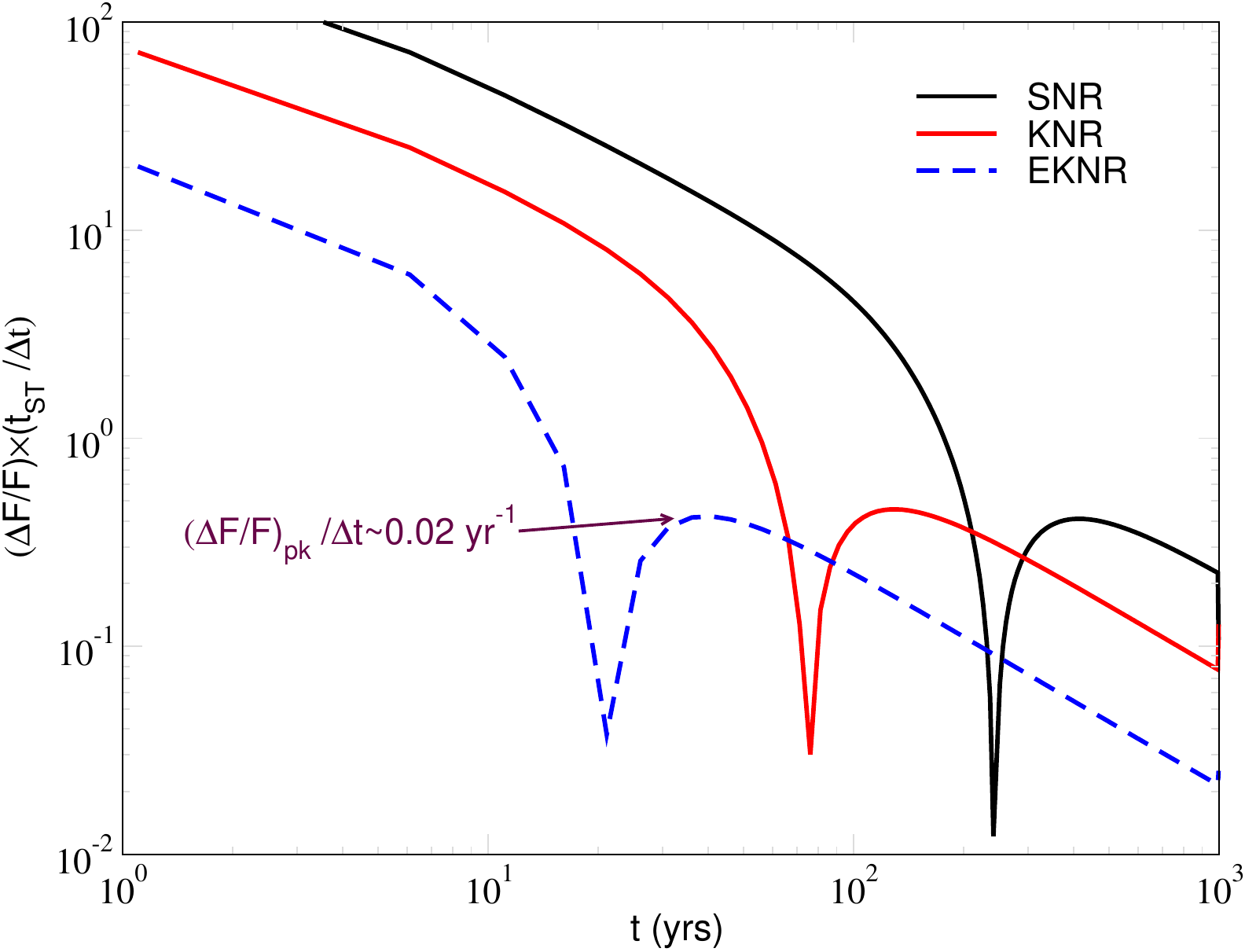}
\end{subfigure}
\caption{Detection prospects for KNRs from their temporal evolution information. Top:$\frac{\Delta F}{F\Delta t}$ for our fiducial cases with $\Delta t=5$ yr. $\frac{{\rm dlog}F_{\rm 3 GHz}}{{\rm d}t}$ approaches $3/t$ and $\sim 1/t$ in the blast wave and ST regimes, respectively. Bottom: $(\frac{\Delta F}{F})\times (\frac{t_{\rm ST}}{\Delta t})$ for the three cases with $\Delta t=5$ yr. We note that $t_{\rm ST}$ is not available to us directly and has to be extracted from the light-curve data. As $(\frac{\Delta F}{F})\times (\frac{t_{\rm ST}}{\Delta t})$ is similar for all cases around the Sedov time, we can estimate $t_{\rm ST}$ from observed quantities, such as $\frac{\Delta F}{F}$ and $\Delta t$ (Eq. \ref{eq:t_ST_estimate}). As an example, we show the case for an EKNR for which $\frac{\Delta F}{F\Delta t}\sim 0.02{\rm yr^{-1}}$ at around the time when it starts to enter deep into the Sedov regime (upper panel).}
\label{fig:F_curvature_1}
\end{figure}

We turn next to the detection prospects for KNRs from the evolution of their light curves. The observed flux $F_{\rm 3 GHz}\propto t^{\alpha}$, where $\alpha$ is a constant, while observing deep within either the coasting or ST phases. The proportionality constant sets the overall magnitude of flux and depends on ejecta parameters and distance. By taking the derivative of the logarithm of the flux, the constant term drops out and we have
\begin{equation}
    \frac{{\rm dlog}F_{\rm 3 GHz}}{{\rm d}t}=\frac{\alpha}{t}+\frac{{\rm d}\alpha}{{\rm d}t}{\rm log t}
.\end{equation}
At $t\ll t_{\rm ST}$ or $t\gg t_{\rm ST}$, the second term drops out with $\alpha=3$ or $-21/20,$ respectively. In fig. \ref{fig:F_curvature_1}, we plot the discretised version of $\frac{|{\rm dlog}(F_{\rm 3 GHz})|}{{\rm d}t}$, assuming observation intervals of $\Delta t=5$ years for our fiducial cases. Far from their respective Sedov time, it is not possible to differentiate the three objects, because the light curves approach the same (featureless) power-law scaling. However, most of the objects are expected to be detected in the Sedov-Taylor regime, because the objects spend most of their time in that domain. 

Assuming we detect an object around $t_{\rm ST}$, we want to determine this timescale from the observed $\frac{\Delta F}{F}$ at intervals of $\Delta t$. From Fig. \ref{fig:F_curvature_1}, we see that, at around $t_{\rm ST}$, $\frac{\Delta F}{F}$ is higher for EKNRs, which is expected given that $t_{\rm ST}$ is smaller than for SNRs, and thus the flux changes faster. As the shape of the light curve is the same in all cases, in dimensionless units they must all be identical. As a visual demonstration, we plot $(\frac{\Delta F}{F})\times (\frac{t_{\rm ST}}{\Delta t})$ for our fiducial cases in the bottom panel of Fig. \ref{fig:F_curvature_1}. We find that $(\frac{\Delta F}{F})\times (\frac{t_{\rm ST}}{\Delta t})$ line up horizontally around their respective $t_{\rm ST}$. As an example, $(\frac{\Delta F}{F})\times (\frac{t_{\rm ST}}{\Delta t})\sim 0.5$ when the light curves are about to enter deep into the ST regime. Therefore, if we detect an object at this epoch, we can obtain an estimate for $t_{\rm ST}$ as,
\begin{equation}
    t_{\rm ST}\sim 0.5\Delta t\left(\frac{\Delta F}{F}\right)_{\rm pk}^{-1},
    \label{eq:t_ST_estimate}
\end{equation}
where $\left(\frac{\Delta F}{F}\right)_{\rm pk}$ corresponds to the value close to the light curve peak.
For EKNRs, $\left(\frac{\Delta F}{F\Delta t}\right)_{\rm pk}\sim 0.02 {\rm yr^{-1}}$. By incorporating this value into Eq. 37, we have $t_{\rm ST}\sim 25$ years. We expect to detect a greater number of KNRs with higher external density
(Sect. \ref{sec:extended_dist}), which have even lower $t_{\rm ST}$. If we find such an object at around its $t_{\rm ST}$, 5-10 years of detailed observations with finer resolution in $\Delta t$ could help to obtain a precise light curve, which would be helpful in distinguishing KNRs from SNRs and other contaminants. As an example, with finer resolution, we can reconstruct the feature that we see around $t_{\rm ST}$. With coarse resolution, we may be unable to find this feature in the data as $t_{\rm ST}\sim 20$ yr for EKNRs.

We should point out that observing such an object can be challenging, as $\Delta F / F \sim 0.1$ over a 5 yr time period can be confused with spatial variation of the surrounding gas density, scintillation, or calibration issues, and so on. As an example, \cite{Mooley2016} used the criteria of $\Delta F / F \gtrsim 0.3$ at S/N$\gtrsim$ 5. We note that the radio afterglow of GW170817 does not show a deviation of more than 30 percent from a power-law spectrum \cite{Mooley2018}. In addition, the calibration error can also be of the order of 5 percent \citep{Ofek2011}.  Therefore, one needs high-S/N events so that the observational uncertainty is $<10$ percent. However, the  fractional change we are discussing here is smooth, whereas fluctuations due to scintillation or density inhomogeneities will likely be stochastic. Furthermore, large sources at close distances ---that is, with $R\gtrsim 10^{19}$cm at  $D<100$Mpc, as in the remnants discussed here--- will be resolved by the scintillation screen (see \citealt{Kumar2024MNRAS} for details) and as such are very weakly affected by scintillation.

\subsection{Chance coincidence of KNR with other radio sources}
As we expect the candidate KNRs to be point sources at distances of $\sim 10$ Mpc (see Sect. \ref{sec:N_remnants}), these can be confused with other radio sources. We compute the probability of a chance coincidence of a candidate source with another radio object.
The source count at $S>40$ $\mu$Jy is given by \citep{Richards2000},
\begin{equation}
    \frac{{\rm dn}}{{\rm d}S}=8.3S^{-2.4}\hspace{0.2cm} {\rm Jy^{-1}Sr^{-1}}
    \label{eq:dn_dS}
.\end{equation}
The source count of VLASS at a threshold of $\sim$ 1 mJy \citep{Gordon2021} matches very well with a six-degree polynomial fit \citep{Hopkins2003}, which differs somewhat from the simple power-law fit at $S>0.1$ mJy. For simplicity, we use the expression above for our estimation. The probability of a chance coincidence is given by \citep{BKD2002,EB2017}
\begin{equation}
    P_{\rm ch}=1-{\rm e}^{-\pi \sigma \theta^2}\approx \pi \sigma \theta^2 ,
\end{equation}
where $\sigma$ is the number of sources per solid angle and $\theta$ can be the angular size of a source such as a galaxy.  From Eq. \ref{eq:dn_dS}, we obtain $n(S)$ as a function of $S$. For a full sky survey, such as VLASS, we obtain $\sigma\sim 2\times 10^{-6}$ arcsec$^{-2}$ at 1 mJy. In order to identify a source as a potential remnant candidate, we need to obtain its distance, and from that we should be able to identify the host galaxy. For a galaxy of $\sim$ 10 kpc  in size at $D\sim 10$ Mpc, the angular size is of the order of arcminutes. Putting everything together with appropriate normalisation, the probability of a chance coincidence is given by
\begin{equation}
    P_{\rm ch}\sim 0.02\left(\frac{S}{\rm 1 mJy}\right)^{-1.4}\left(\frac{\theta}{\rm arcmin}\right)^2
.\end{equation}
For a galaxy of 10 kpc in size, we find that the probability of a chance coincidence can be of the order of a few percent.  This might be an issue where the overall number of candidate objects is a few, but they may be distinguishable using spectral (Sect. \ref{sec:electron_properties}) and light-curve information. 

The EKNRs are expected to be much brighter with flux of $\sim 100$ mJy (Fig. \ref{fig:F_3GHz}) at a distance of 10 Mpc. In that case, the probability of a chance coincidence is given by
\begin{equation}
    P_{\rm ch}\sim 3.2\times 10^{-5}\left(\frac{S}{\rm 100 mJy}\right)^{-1.4}\left(\frac{\theta}{\rm arcmin}\right)^2
.\end{equation}
In these expressions, we have considered all objects that are seen in the radio sky. However, we can use the flux variability information to reduce the chance coincidence even further. The authors in \cite{Becker2010}, \cite{Bannister2011} and  \cite{Croft2010} show that the number of variable sources on a timescale of 7-22 yr is of the order of a few percent. Using VLA and three epochs spanning 15 yr, \cite{Becker2010} show that the number of extragalactic sources with a variability of greater than 2 is less than 10 percent within the flux-density range of 1-100 mJy. Therefore, using the variability information we can increase the confidence of our detection even further.

\subsection{Distinguishing between KNRs and SNRs with flux information}
\label{subsec:KNR_vs_SNR}
In the above sections, we discuss the importance of temporal measures such as the Sedov time in order to distinguish a KNR from a SNR. The measurement of flux at a given epoch may still be used for that purpose provided we know the density of the galaxy in which the KNR is located. The reader may notice that the peak flux for both SNRs and KNRs, assuming fiducial parameters, is the same in Fig. \ref{fig:F_3GHz}, which follows from Eq. \ref{eq:F@tST}. This may appear misleading, as the assumed external density encountered by KNRs is ten times smaller than that for SNRs. For a given density, the peak flux of a KNR is expected to be many times higher than that of a SNR, which becomes even more drastic in the case of EKNRs. Also, the spectral flux at different frequency bands (Fig. \ref{fig:nu_Lnu}) will also be higher in the case of KNRs. In addition, the SNRs are expected to show distinctive line features \citep{WW1986,BHBL2007} as well as morphologies \citep{LRBHJP2009} in X-ray, which allows us to distinguish Type Ia from core-collapse supernovae. If we find a candidate object in an elliptical galaxy, we can almost certainly rule out a core-collapse supernova as the progenitor. Then, the non-detection of expected line features or morphology of Type Ia supernovae may help us in distinguishing a candidate KNR. Furthermore, as explained above, the peak luminosity for a KNR will be significantly larger. If we detect an object with  $L_{3{\rm GHz}}\gtrsim 10^{27}$ erg s$^{-1}$Hz$^{-1}$ associated with a remnant in an elliptical galaxy, this would be a very strong indication that the remnant is a KNR (or EKNR).

\section{Discussion and conclusion}
\label{sec:conclusions}
We have solved for the evolution of kilonova remnants (KNRs) and compare their light curves and spectral properties with supernova remnants (SNRs). While the physics is the same in both cases, we expect differences in quantitative properties due to different fiducial parameters, such as lower ejecta mass and lower external density in the KNR case. Spectral signatures such as the synchrotron cutoff, the correlation between radio and X-ray flux, and the transition from the synchrotron- to the SSC-dominated regime  are features that can be used to distinguish KNRs from SNRs. As an example, around the peak of their flux, the synchrotron cutoff is of the order of 10 keV, 100 keV, and MeV for SNRs, KNRs, and EKNRs, respectively. We also note that despite the fact that most remnants are much older than the time at which their light curve peaks, due to the strong decrease in the luminosity with age during the decelerating phase, the observable sample is dominated by events observed close to their peak flux. However, as ${\rm dN/dlog(t)} \sim t^{-1/2}$, one in three remnants will be approximately ten times older than $t_{\rm ST}$, which is not insignificant. 

For fiducial parameters ($E_{51}=1$, $M_{\rm ej,\odot}$= 0.1, $n_0$= 0.1), we expect to detect KNRs in the near future using existing telescopes, such as VLASS. Due to the strong dependence of the observed flux on ejecta energy, extreme KNe ($E_{51}=10$) are much easier to detect if they constitute a sizeable portion of the total KN population. As there have been no reported detections of EKNRs in the literature, we are able to put moderate constraints on their abundance relative to regular KNRs, with $f_{\rm EKN}\lesssim 0.3$ assuming fiducial parameters. This constraint depends sensitively on the ejecta mass and external density distribution,  as well as the neutron star merger rate, which is somewhat uncertain. We should be able to detect such objects or put strong constraints on their occurrence rate with ongoing surveys, such as VLASS and VAST, or future surveys with DSA-2000 and SKA \citep{DSA2000,SKA}.

The Sedov time for a EKNR is $\sim$20 years, which is at least an order of magnitude less than for a SNR. Depending on the external density, the Sedov time can be even shorter. If the remnants are detected deep in the Sedov-Taylor regime, we will not be able to easily distinguish KNRs from SNRs. However, if we do detect a remnant around its Sedov time, it may be possible to obtain a good estimate of $t_{\rm ST}$ with a few data points taken with a time interval of several years. A combination of estimated $t_{\rm ST}$  and corresponding spectral properties can provide strong evidence for the existence of a KNR. Also, if we find a candidate object in an elliptical galaxy, we can distinguish a KNR from a type Ia SNR using its brightness information or using line features and morphology in X-rays.

From the only robust KN identification to date, which is associated with the GW-detected NS merger GW170817, we can infer the ejecta mass and velocity for a given electron mass fraction $Y_e$. Different choices of $Y_e$ can result in different interpretations of the properties of the ejecta \citep{2019LRR....23....1M}. If we are able to infer ejecta mass and velocity from a remnant, we should be able to break the degeneracy and can independently infer $Y_e$; this will have implications in terms of $r$-process nucleosynthesis and help us understand how much radioactive material was synthesised during the formation of the KN.

\begin{acknowledgements}
SKA is supported by ARCO fellowship. PB's work was supported by a grant (no. 2020747) from the United States-Israel Binational Science Foundation (BSF), Jerusalem, Israel, by a grant (no. 1649/23) from the Israel Science Foundation and by a grant (no. 80NSSC 24K0770) from the NASA astrophysics theory program. KH's work was supported by JSPS KAKENHI Grant-in-Aid for Scientific Research JP23H01169, JP23H04900, and JST FOREST Program JPMJFR2136.
\end{acknowledgements}

\bibliographystyle{aa}
\bibliography{main}

\begin{appendix}
\section{Relativistic equations for dynamical properties of kilonovae}
\label{subsec:dynamics_relativistic}
In this section we provide fully relativistic equations for the evolution of ejecta. For this work, we find that relativistic corrections are mildly important for some extreme parameter cases. For such cases, we directly solve these equations numerically. The dynamical equation for the ejecta can be written as
\begin{equation}
    E=(\Gamma-1)M_{\rm ej}{\rm c^2}+\frac{4\pi}{3}n_{\rm ISM}R_{\rm ej}^3m_{\rm p}{\rm c^2}\Gamma(\Gamma-1),
\end{equation}
where $E$ is the kinetic energy, $\Gamma$ is the bulk Lorentz factor and $R_{\rm ej}$ is the size of ejecta in observer frame. The ejecta size is related to the observed time as $R_{\rm ej}= (1+\beta_{\rm ej})\Gamma^2 \beta_{\rm ej}c t$. The rest of the symbols have the same interpretation as in Sect. \ref{sec:dynamics_newtonian}. In the coasting phase, $\Gamma=1.0+\left(\frac{E_{51}}{M_{\rm ej,\odot}}\right)\times 5\times 10^{-4}$.

Assuming a fraction $\epsilon_e$ of the shocked energy is transferred to the electrons, by the same procedure as in Sect. \ref{sec:electron_properties}, we obtain
\begin{equation}
    \gamma_m-1= \frac{p-2}{p-1}\frac{\epsilon_e}{\xi_e}\frac{m_{\rm p}}{m_{\rm e}}(\Gamma-1)
.\end{equation}
The comoving magnetic energy density is given by
\begin{equation}
    B=(32\pi\Gamma(\Gamma-1)n_{\rm ISM}m_{\rm p}{\rm c^2}\epsilon_B)^{1/2}.
\end{equation}
The synchrotron energy loss rate per electron in observer frame is given by
\begin{equation}
    \frac{{\rm d}E_e}{{\rm d}t}=\frac{4}{3}\sigma_{\rm T}{\rm c}\Gamma^2\gamma_m^2\beta_m^2 B^2,
\end{equation}
where $E_e$ is the electron's energy. The characteristic frequency in observer frame is given by
\begin{equation}
    \nu_m=\Gamma\gamma^2_m\nu_B,
\end{equation}
The power in observer frame is then
\begin{equation}
    P_m=3.8\times 10^{-22}\Gamma\left(\frac{B}{\rm 1 G}\right)      \mbox{ erg s}^{-1}\mbox{Hz}^{-1},
\end{equation}
assuming $\beta_m\rightarrow 1$. The calculation of luminosity and flux follows the same way as in Sect. \ref{sec:synchrotron}.
\section{Cooling due to synchrotron}
\label{subsec:nu_c}
The Lorentz factor $\gamma_{\rm c,sync}$ for which the electrons cool within the timescale of expansion of the ejecta, i.e. the dynamical time, can be evaluated as,   $ \gamma_{\rm c,sync} m_{\rm e}{\rm c^2}=\frac{4}{3}\sigma_{\rm T}\gamma_{\rm c,sync}^2\frac{B^2}{8\pi}{\rm c}t$,
or 
\begin{equation}
\gamma_{\rm c,sync}=\frac{6\pi m_{\rm e}{\rm c}}{\sigma_{\rm T}B^2t} 
\label{eq:gamma_c_sync}
\end{equation}
for $\gamma_{\rm c,sync}>>1$ which can be simplified to, $\gamma_{\rm c,sync}=\frac{24.5}{B^2t_{\rm yr}}$.
The cooling frequency is given by, $\nu_{\rm c,sync}=\gamma_{\rm c,sync}^2\nu_B$. 
In the coasting phase, $B$ is constant, hence, $\nu_{\rm c,sync}$ goes like $t_{\rm yr}^{-2}$. Substituting the expression for $B$ is ST regime, we have
\begin{equation}
\begin{aligned}
    \nu_{\rm c,sync}\sim 1.7\times 10^{18} \epsilon_{B,-2}^{-3/2}E_{51}^{-3/2}n_0^{-3/2}M_{\rm ej,\odot}^{3/2}t_{\rm yr}^{-2}\hspace{0.2cm} {\rm Hz}\hspace{0.2cm} {(\rm coasting\hspace{0.1cm}phase)} \\
    \nu_{\rm c,sync}\sim 6.12\times 10^{14} \epsilon_{B,-2}^{-3/2}E_{51}^{-3/5}n_0^{-9/10}t_{\rm yr}^{-1/5}\hspace{0.2cm} {\rm Hz}\hspace{0.2cm} {(\rm ST\hspace{0.1cm}phase)}
\end{aligned}    
.\end{equation}

\section{Synchrotron self-absorption frequency ($\nu_{\rm sa}$)}
\label{subsec:nu_sa}
In the optically thin regime, the power per Hz at $\nu_m$ per steradian is given by
\begin{equation}
    P_{\nu_m}=\frac{3.8\times 10^{-22}}{4\pi}\left(\frac{B}{\rm 1 G}\right) \hspace{0.5cm} \mbox{ erg s}^{-1} \mbox{Hz}^{-1} \mbox{sr}^{-1}
.\end{equation}
Assuming $\nu_{\rm sa}>\nu_m$, the power at $\nu_{\rm sa}$ is given by,
 $   P(\nu_{\rm sa})=P_{\nu_m}\left(\frac{\nu_{\rm sa}}{\nu_m}\right)^{-(p-1)/2}$.
The intensity at $\nu_{\rm sa}$ is given by
\begin{equation}
    I_{\nu}=0.3\times 10^{-3}B\left(\frac{\nu_{\rm sa}}{\nu_m}\right)^{-(p-1)/2}\left(\frac{R}{10^{19} cm}\right)n_{0}\xi_e \hspace{0.5cm} \mbox{ erg  cm}^{-2} \mbox{Hz}^{-1}\mbox{sr}^{-1}
.\end{equation}
The intensity of a blackbody is given by
\begin{equation}
    I_{\nu}=\frac{2k_BT}{2.7}\frac{\nu_{\rm sa}^2}{\rm c^2},
\end{equation}
where $k_BT\sim \gamma_{\rm sa} m_e{\rm c^2}$ with $\gamma_{\rm sa}=\sqrt{\nu_{\rm sa}/\nu_B}$. Equating the two and using $p=2.5$, the expression for $\nu_{\rm sa}$ is given by
\begin{equation}
    \nu_{\rm sa}^{3.25}=15\times 10^{31}B^{2.25}R_{19}n_{0}\xi_e, \hspace{0.5cm} 
\end{equation}
 or $\nu_{\rm sa}\sim 8\times 10^9 (B^{2.25}R_{19}n_0\xi_e)^{1/3.25}$\hspace{0.2cm} Hz. Substituting the expressions for the physical parameters, we have
 \begin{equation}
     \begin{aligned}
     \nu_{\rm sa}\sim 3\times 10^6 {\bar{\epsilon}_{e,-1}}^{4/13}\epsilon_{B,-2}^{9/26}E_{51}^{21/26}M_{\rm ej,\odot}^{-21/26}n_0^{17/26}t_{\rm yr}^{0.31}\hspace{0.1cm}{\rm Hz}\hspace{0.2cm} ({\rm coasting\hspace{0.1cm} phase}) \\
     \nu_{\rm sa}\sim 4\times 10^8 {\bar{\epsilon}_{e,-1}}^{4/13}\epsilon_{B,-2}^{9/26}E_{51}^{21/65}n_0^{43/130}t_{\rm yr}^{-43/65}\hspace{0.1cm}{\rm Hz}\hspace{1.4cm} ({\rm ST\hspace{0.05cm}phase}) 
     \end{aligned}
 .\end{equation}

\section{Computation of SSC flux}
\label{subsec:ssc_flux}
The number flux of inverse Compton photons is given by \citep{BG1970,SE2001}
\begin{equation}
    f_{\nu,{\rm IC}}=\sigma_TR_{\rm ej}\int_{\gamma_m}^{\infty}d\gamma_e\frac{{\rm d}N_e}{{\rm d}\gamma_e}\int_0^{\infty}{\rm d}\nu_s g(x)f_{\nu_s}(x)
,\end{equation}
where $x=\frac{\nu}{4\gamma_e^2\nu_s}$, $g(x)=1+x-2x^2+2x{\rm log}(x)$ in Thomson approximation and $f_{\nu_s}=\frac{L_{\nu_s}}{h\nu_s}$ where $L_{\nu_s}$ is obtained in Sect. \ref{sec:synchrotron}. Technically, the flux on both sides of the equation is at the location of shocked gas but they can be transformed to the observed flux with the same factor of $\frac{R_{\rm ej}^2}{D^2}$ which cancels out. The factor $g(x)$ has support only over $0\lesssim x \lesssim 1$ which effectively cuts off the $\nu_s$ integral for a given $\nu$. Expanding the definition of the quantities, we have
\begin{equation}
    \frac{{\rm d}N_{{\gamma},\rm IC}}{{\rm d}t{\rm d}\nu}=\sigma_TR_{\rm ej}\int_{\gamma_m}^{\infty}d\gamma_e\frac{{\rm d}N_e}{{\rm d}\gamma_e}\int_0^{\infty}{\rm d}\nu_s g(x)f_{\nu_s}(x)\frac{1}{4\gamma_e^2\nu_s}
.\end{equation}
We solve this equation numerically and obtain the number flux which can be converted to $L_{\nu}$ by multiplying $h\nu$. We compare synchrotron, SSC and bremsstrahlung flux for our fiducial cases in Fig. \ref{fig:nu_Lnu}.

\section{Bremsstrahlung emission}
\label{subsec:bremsstrahlung}
The thermal energy of electron is related to the shock energy via electron kinetic energy. Therefore, the temperature of electron is given by
\begin{equation}
  \bar{\epsilon}_e\frac{1}{2}m_{\rm p}\beta_{\rm ej}^2{\rm c^2}=\frac{3}{2} k_B T_{\rm e}, \\
  \label{eq:T_fs_eq}
\end{equation}
which after substitution of $\beta_{\rm ej}$ becomes
\begin{equation}
\begin{aligned}
    T_{\rm e}= 3\times 10^{8}\bar{\epsilon}_{e,-1}\frac{E_{51}}{M_{\rm ej,\odot}}\hspace{0.1cm} {\rm K}\hspace{1.5cm} {(\rm coasting\hspace{0.1cm} phase)} \\
    T_{\rm e}=0.88\times 10^{11}\bar{\epsilon}_{e,-1}E_{\rm 51}^{2/5}n_0^{-2/5}t_{\rm yr}^{-6/5}\hspace{0.1cm} {\rm K}\hspace{0.2cm} {(\rm ST\hspace{0.1cm} phase)}
    \label{eq:T_e}
    \end{aligned}
.\end{equation}
The spectrum has a cutoff frequency $\nu_{\rm b,co}\sim \frac{k_B T_{\rm e}}{h}$.
The total emitted power per unit volume per frequency is given by
\begin{equation}
    \frac{{\rm d}W}{{\rm d}V{\rm d}t{\rm d}\nu}=6.8\times 10^{-38}Z^2n_i n_eT^{-1/2}{\rm e}^{-h\nu/kT}g_{ff} \hspace{0.5cm} \mbox{erg cm}^{-3}\mbox{ s}^{-1} \mbox{Hz}^{-1}
.\end{equation}
Assuming $g_{ff}=1$ and $h\nu\lesssim k_BT$, the luminosity per frequency is given by
\begin{equation}
\begin{aligned}
    L_{\nu}=0.42\times 10^{9} n_0^2  \left(\frac{E_{51}}{M_{\rm ej,\odot}}\right)t_{\rm yr}^3\hspace{0.2cm} \mbox{ erg s}^{-1}\mbox{Hz}^{-1}\hspace{0.1cm} ({\rm coasting\hspace{0.05cm}phase}) \\ 
    L_{\nu}=1.2\times 10^{12}E_{\rm 51}^{2/5}n_0^{8/5}t_{\rm yr}^{9/5}\hspace{0.5cm} \mbox{ erg s}^{-1} Hz^{-1}\hspace{0.1cm} ({\rm ST\hspace{0.05cm}phase})
    \end{aligned}
.\end{equation}
The corresponding flux is given by
\begin{equation}
    \begin{aligned}
    F_{\nu}=3\times 10^{-18}\frac{ n_0^2  \left(\frac{E_{51}}{M_{\rm ej,\odot}}\right)t_{\rm yr}^3}{D_{10}^2}\hspace{0.2cm} \rm mJy\hspace{0.1cm} ({\rm coasting\hspace{0.05cm}phase}) \\ 
    F_{\nu}= 10^{-14}\frac{E_{\rm 51}^{2/5}n_0^{8/5}t_{\rm yr}^{9/5}}{D_{\rm 10}^2}\hspace{0.2cm} \rm mJy\hspace{0.6cm} ({\rm ST\hspace{0.05cm}phase})
    \end{aligned}
    .\end{equation}

\section{Number of SNe in the Milky Way}
\label{sec:R_SN}
In this work, we assume the volumetric rate of formation of supernova remnants to be 300 times greater than kilonova. Using this, the number density of SNRs $\sim 10^{5}$ Gpc$^{-3}t_{\rm yr}$ where $t_{\rm yr}$ is the age of SNRs and we have assumed that observed flux is not the limiting factor. We will compute the number of SNRs in the Milky Way and compare it with the number of detected objects which is $\sim 300$ in \cite{Green2019}  as a consistency check. This is a revised catalogue of Galactic SNRs and for each SNR, Galactic coordinates, angular size, flux density, type of SNR and spectral index are provided. We find that most of the sources in the catalogue, have fluxes of the order of few Jy at $\nu\sim {\rm GHz}$. Using the fiducial parameters, the flux in the ST regime in our model is given by (Eq. \ref{eq:flux_nu})
\begin{equation}
    F_{\nu}\sim 8.85\times 10^2 \frac{t_{\rm yr}^{-21/20}}{D_{\rm 10}^2}\left(\frac{\nu}{\rm 3\hspace{0.1cm}GHz}\right)^{-3/4}\hspace{0.2cm} {\rm mJy} 
.\end{equation}
 Assuming all SN are located at 10 kpc from us, $F_{\rm GHz}\sim 2\times 10^6t_{\rm yr}^{-1}$ Jy. Therefore, most SN in the catalogue are $10^5-10^6$ year old. To convert the volumetric formation rate to galactic rate, we use the number density of MW type galaxy $\sim 0.01$ Mpc$^{-3}$ \citep{HB2018}. Combining these and assuming all remnants to be of the order of $10^5$ years old, the expected SNRs in our galaxy turns out to be $\sim \frac{10^5}{\rm 0.01 Mpc^{-3}}\times {\rm Gpc^{-3}}\times 10^5\approx 1000$ which gives roughly similar numbers to those in the catalogue.
\end{appendix}

\end{document}